\documentclass[showpacs,aps,pre,twocolumn]{revtex4}

\usepackage{amsmath}
\usepackage{amssymb}
\usepackage{amsbsy}
\usepackage{makeidx}
\usepackage{epsf}
\usepackage{graphics}
\usepackage{graphicx}
\usepackage{amsfonts}
\usepackage{subfigure}

\newcommand{\Ord}{{\mathcal O}}

\begin{document}
\title{Discrete solitons in electromechanical resonators}
\author{M. Syafwan}
\altaffiliation[On leave from ]{Department of Mathematics,
Faculty of Mathematics and Natural Sciences, Andalas University,
Limau Manis, Padang, Indonesia 25163}
\author{H. Susanto}
\author{S. M. Cox}
\affiliation{School of Mathematical Sciences, University of Nottingham,
University Park, Nottingham, NG7 2RD, UK}


\begin{abstract}

We consider a parametrically driven Klein--Gordon system describing
micro- and nano-devices, with integrated electrical and mechanical
functionality. Using a multiscale expansion method we reduce the system
to a discrete nonlinear Schr\"odinger equation. Analytical and numerical
calculations are performed to determine the existence and stability of
fundamental bright and dark discrete solitons admitted by the
Klein--Gordon system through the discrete Schr\"odinger equation. We
show that a parametric driving can not only destabilize onsite bright
solitons, but also stabilize intersite bright discrete solitons and
onsite and intersite dark solitons. Most importantly, we show that there
is a range of values of the driving coefficient for which dark solitons
are stable, for any value of the coupling constant, i.e.\ oscillatory
instabilities are totally suppressed. Stability windows of all the
fundamental solitons are presented and approximations to the onset of
instability are derived using perturbation theory, with accompanying
numerical results. Numerical integrations of the Klein--Gordon equation
are performed, confirming the relevance of our analysis.

\end{abstract}

\maketitle

\section{Introduction}

Current advances in the fabrication and control of electromechanical
systems on a micro and nanoscale bring many technological
promises~\cite{vent04}. These include efficient and highly sensitive
sensors to detect stresses, vibrations and forces at the atomic level,
to detect chemical signals, and to perform signal
processing~\cite{clel02}. As a particular example, a
nanoelectromechanical system (NEMS) can detect the mass of a single atom,
due to its own very small mass~\cite{rouk01,clel98}.

On a fundamental level, NEMS with high frequency will allow research on
quantum mechanical effects. This is because NEMS, as a miniaturization
of microelectromechanical systems (MEMS), can contain a macroscopic
number of atoms, yet still require quantum mechanics for their proper
description. Thus, NEMS can be considered as a natural playground for a
study of mechanical systems at the quantum limit and
quantum-to-classical transitions (see, e.g., Ref.~\onlinecite{katz08}
and references therein).

Typically, nanoelectromechanical devices comprise an electronic device
coupled to an extremely high frequency nanoresonator. A large number of
arrays of MEMS and NEMS resonators have recently been fabricated
experimentally (see, e.g., Ref.~\cite{buks02}). One direction of research
in the study of such arrays has focused on intrinsic localized modes
(ILMs) or discrete breathers. ILMs can be present due to parametric
instabilities in an array of oscillators~\cite{wiersig}. ILMs in driven
arrays of MEMS have been observed
experimentally~\cite{sato06,sato07_1,sato07_2}.


Motivated by a recent experiment of Buks and Roukes~\cite{buks02} that
succeeded in fabricating and exciting an array of MEMS and measuring
oscillations of the resonators, here we consider the equation~\cite{keni09}
 \begin{equation}
\ddot{\varphi}_n=
D\Delta_2\varphi_n-[1-H\cos(2\omega_pt)]\varphi_n\pm \varphi_n^3,
\label{eq1}
 \end{equation}
which governs the oscillation amplitude of such an array. Equation
(\ref{eq1}) is a simplified model of that discussed in
Ref.~\onlinecite{lifs03}, subject to an assumption that the
piezoelectric parametric drive is applied directly to each
oscillator~\cite{msm07}. The variable $\varphi_n$ represents the
oscillation amplitude of the $n$th oscillator from its equilibrium
position, $D$ is a dc electrostatic nearest-neighbor coupling term, $H$
is a small ac component with frequency $2\omega_p$ responsible for the
parametric driving,
$\Delta_2\varphi_n=\varphi_{n+1}-2\varphi_n+\varphi_{n-1}$ is the
discrete Laplacian, the dot denotes the derivative with respect to $t$,
and the `plus' and `minus' signs of the cubic term correspond to a
`softening' and `stiffening' nonlinearity, respectively. Here, we assume
ideal oscillators, so there is no damping present. The creation,
stability, and interactions of ILMs in (\ref{eq1}) with low damping and
in the strong-coupling limit, have been investigated in
Ref.~\onlinecite{keni09}. Here, we extend that study to the case of
small coupling parameter $D$.

In performing our analysis of the governing equation (\ref{eq1}), we introduce
a small parameter $\epsilon\ll1$, and assume that the following scalings hold:
 \[
D=\epsilon^23C,\quad
H=\mp\epsilon^23\gamma,\quad
\omega_p=1\mp\epsilon^23\Lambda/2.
 \]
We then expand each $\varphi_n$ in powers of $\epsilon$, with the leading-order
term being of the form
 \begin{eqnarray}
\varphi_n\sim
\epsilon\left(\psi_n(T_2,T_3,\dots)e^{-iT_0}+
\overline{\psi}_n(T_2,T_3,\dots)e^{iT_0}\right),
\label{expand}
 \end{eqnarray}
where $T_n=\epsilon^n T$. Then the terms at
$\Ord(\epsilon^3e^{-iT_0})$ in (\ref{eq1}) yield the
following equation for $\psi_n$ (see Refs.~\onlinecite{kivs93,remo86} for a related
reduction method):
 \begin{equation}
-2i\dot{\psi}_n=
3C\Delta_2\psi_n\mp3\gamma\overline{\psi}_ne^{\pm i3\Lambda T_2}
\pm3|\psi_n|^2\psi_n,
\label{gov1}
 \end{equation}
where the dot now denotes the derivative with respect to $T_2$.
Correction terms in Eq.~(\ref{expand}) are of order
$\Ord(\epsilon e^{\pm i(k+1)T_0},\epsilon^3e^{\pm
i(k-1)T_0}),\,k\in\mathbb{Z}^+$. A justification of this rotating wave type approximation
can be obtained in, e.g., Ref.\ \onlinecite{kose02}.

Writing $\psi_n(T_2)=\phi_n(T_2)e^{\pm i3\Lambda/2 T_2}$, we find that
Eq.~(\ref{gov1}) becomes
 \begin{equation}
-\frac23i\dot{\phi}_n=
C\Delta_2\phi_n\mp\Lambda\phi_n\mp\gamma\overline{\phi}_n\pm|\phi_n|^2\phi_n.
 \end{equation}
Then, taking $T_2=\frac23\hat{T}$, we find that the equation above becomes
the parametric driven discrete nonlinear Schr\"odinger (DNLS) equation
 \begin{equation}
i\dot{\phi}_n=
-C\Delta_2\phi_n\pm\Lambda\phi_n\pm\gamma\overline{\phi}_n\mp|\phi_n|^2\phi_n;
\label{gov}
 \end{equation}
here the dot denotes the derivative with respect to $\hat{T}$. The
softening and stiffening nonlinearity of (\ref{eq1}) correspond, respectively,
to the so-called focusing and defocusing nonlinearity in the DNLS~(\ref{gov}).

In the absence of parametric driving, i.e., for $\gamma=0$,
Eq.~(\ref{gov}) is known to admit bright and dark solitons in the system
with focusing and defocusing nonlinearity, respectively. Discrete bright
solitons in such a system have been discussed before, e.g.~in
Refs.~\onlinecite{henn99,alfi04,peli05}, where it was shown that
one-excited-site (onsite) solitons are stable and two-excited-site
(intersite) solitons are unstable, for any coupling constant $C$.
Undriven discrete dark solitons have also been
examined~\cite{fitr07,joha99,kivs94,susa05,peli08}; it is known that
intersite dark solitons are always unstable, for any $C$, and onsite
solitons are stable only in a small window in $C$. Furthermore, an
onsite dark soliton is unstable due to the presence of a quartet of
complex eigenvalues, i.e., it suffers oscillatory instability.

The parametrically driven DNLS (\ref{gov}) with a focusing nonlinearity
and finite $C$ has been considered briefly in Ref.~\onlinecite{susa06},
where it was shown that an onsite bright discrete soliton can be
destabilized by parametric driving. Localized excitations of the
continuous limit of the parametrically driven DNLS, i.e.~(\ref{gov})
with $C\to\infty$, have been considered by Barashenkov et al.\ in a
different context of applications
\cite{bara96,bara01,bara03,bara03_2,bara05,bara07,bara07_2}. The same
equation also applies to the study of Bose--Einstein condensates,
describing the so-called long bosonic Josephson junctions
\cite{kukl05,kukl06}.

In this paper, we consider (\ref{eq1}) with either softening or
stiffening nonlinearities, which admit bright or dark discrete solitons,
respectively. The existence and stability of the fundamental onsite and
intersite excitations are discussed through the reduced equation
(\ref{gov}). Eq.~(\ref{gov}) and a corresponding eigenvalue problem are
solved numerically for a range of values of the coupling and driving
constants, $C$ and $\gamma$, giving stability windows in the
$(C,\gamma)$ plane. Analytical approximations to the boundaries of the
numerically obtained stability windows are determined through a
perturbation analysis for small $C$. From this analysis, we show,
complementing the result of Ref.~\onlinecite{susa06}, that parametric
driving can stabilize intersite discrete bright solitons. We also show
that parametric driving can even stabilize dark solitons, for any
coupling constant $C$. These findings, which are obtained from the
reduced equation (\ref{gov}), are then confirmed by direct numerical
integrations of the original governing equation (\ref{eq1}).

The present paper is organized as follows. In Sec.~II we present the
existence and stability analysis of onsite and intersite bright
solitons. Analysis of dark solitons is presented in Sec.~III.
Confirmation of this analysis, through numerical simulations of the
Klein--Gordon system (\ref{eq1}), is given in Sec.~IV. Finally, we give
conclusions in Sec.~V.

\section{Bright solitons in the focusing DNLS}

In this section we first consider the existence and stability of bright
solitons in the focusing DNLS equation. For a static solution of
\eqref{gov} of the form $\phi_n=u_n$, where $u_n$ is real-valued and
time-independent, it follows that
 \begin{equation}
-C\Delta_2u_n-u_n^3+\Lambda u_n+\gamma u_n=0.
\label{schr2}
 \end{equation}

Once such discrete solitary-wave solutions of \eqref{gov} have been
found, their linear stability is determined by solving a corresponding
eigenvalue problem. To do so, we introduce the linearization ansatz
 \[
\phi_n=u_n+ \delta \epsilon_n,
 \]
where $\delta\ll1$,
and substitute this into \eqref{gov}, to yield the following linearized
equation at $\Ord(\delta)$:
 \begin{equation}
i\dot{\epsilon}_n=
-C\Delta_2\epsilon_n-2|u_n|^2\epsilon_n-
u_n^2\overline{\epsilon}_n+\Lambda\epsilon_n+\gamma\overline{\epsilon}_n.
\label{lin}
 \end{equation}
Writing $\epsilon_n(t)=\eta_n+i\xi_n$, we then obtain from
Eq.~\eqref{lin} the eigenvalue problem
 \begin{eqnarray}
\left(\begin{array}{cc}
\dot{\eta}_n\\
\dot{\xi}_n
\end{array}\right)=
{\cal H}
\left(\begin{array}{cc}
{\eta}_n\\
{\xi}_n
\end{array}\right),
\label{lin2}
 \end{eqnarray}
where
 \[
{\cal H}=
\left(
\begin{array}{cc}
0 & {\cal L}_+(C) \\
-{\cal L}_-(C) & 0
\end{array}
\right)
 \]
and the operators ${\cal L}_-(C)$ and ${\cal L}_+(C)$ are defined by
 \begin{eqnarray*}
\mathcal{L}_-(C)&\equiv&-C\Delta_2-(3u_n^2-\Lambda-\gamma),\\
\mathcal{L}_+(C)&\equiv&-C\Delta_2-(u_n^2-\Lambda+\gamma).
 \end{eqnarray*}
The stability of the solution $u_n$ is then determined by the
eigenvalues of \eqref{lin2}. If we denote these eigenvalues by
$i\omega$, then the solution $u_n$ is stable only when Im$(\omega)=0$
for all eigenvalues $\omega$.

We note that, because \eqref{lin2} is linear, we may eliminate one of
the eigenvectors, for instance $\xi_n$, to obtain an alternative
expression of the eigenvalue problem in the form
 \begin{equation}
{\cal L}_+(C){\cal L}_-(C)\eta_n=\omega^2\eta_n\equiv\Omega\eta_n.
\label{evp}
 \end{equation}
In view of the relation $\Omega=\omega^2$, it follows that a soliton is
unstable if it has an eigenvalue with either $\Omega<0$ or
Im$(\Omega)\neq0$.

\subsection{Analytical calculations}

Analytical calculations of the existence and stability of discrete
solitons can be carried out for small coupling constant $C$, using a
perturbation analysis. This analysis exploits the exact solutions of
\eqref{schr2} in the uncoupled limit $C=0$, which we denote by
$u_n=u_n^{(0)}$, in which each $u_n^{(0)}$ must take one of the three
values given by
 \begin{equation}
0,\,\pm\sqrt{\Lambda+\gamma}.
\label{un0}
 \end{equation}
Solutions of \eqref{schr2} for small $C$ can then be calculated
analytically by writing
 \[
u_n=u_n^{(0)}+Cu_n^{(1)} + C^2 u_n^{(2)} + \cdots.
 \]

In studying the stability problem, it is natural to also expand the
eigenvector $\eta_n$ and the eigenvalue $\Omega$ in powers of $C$, as
 \[
\eta_n=\eta_n^{(0)}+C\eta_n^{(1)}+\Ord(C^2),\quad
\Omega=\Omega^{(0)}+C\Omega^{(1)}+\Ord(C^2).
 \]

Upon substituting this expansion into Eq.~\eqref{evp} and identifying
coefficients of successive powers of the small parameter $C$, we
obtain from the equations at $\Ord(1)$ and $\Ord(C)$ the
results
 \begin{eqnarray}
\displaystyle
\left[{\cal L}_+(0){\cal L}_-(0)-\Omega^{(0)}\right]\eta_n^{(0)}&=&0,
\label{o0}\\
\displaystyle
\left[{\cal L}_+(0){\cal L}_-(0)-\Omega^{(0)}\right]\eta_n^{(1)}&=&
f_n\eta_n^{(0)},
\label{o1}
 \end{eqnarray}
where
 \begin{equation}
\displaystyle
f_n=-(\Delta_2+2u^{(0)}_nu^{(1)}_n){\cal L}_-(0)-
{\cal L}_+(0)(\Delta_2+6u_n^{(0)}u_n^{(1)})+\Omega^{(1)}.
\label{f}
\end{equation}

In the uncoupled limit, $C=0$, the eigenvalue problem is thus simplified to
 \begin{eqnarray}
\Omega^{(0)}={\cal L}_+(0){\cal L}_-(0),
 \end{eqnarray}
from which we conclude that there are two possible eigenvalues, given by
 \[
\Omega^{(0)}_C=\Lambda^2-\gamma^2,\quad
\Omega^{(0)}_E=4(\Lambda+\gamma)\gamma,
\]
which correspond, respectively, to the solutions $u_n^{(0)}=0$ (for all $n$)
and $u_n^{(0)}=\pm\sqrt{\Lambda+\gamma}$ (for all $n$).

\begin{figure}[tbhp]
\centering
\includegraphics[width=6cm,clip=]{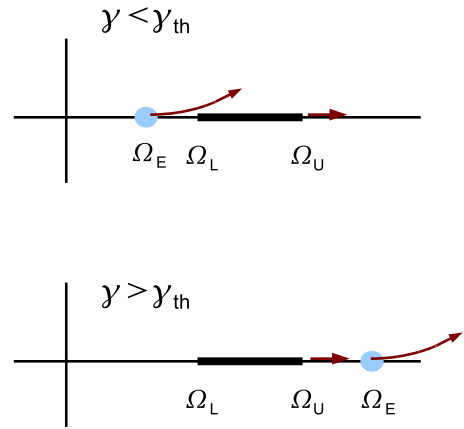}

\caption{A sketch of the dynamics of the eigenvalues and the continuous
spectrum of a stable onsite bright soliton in the
(Re($\Omega$),Im($\Omega$)) plane. The arrows indicate the direction of
movement as the coupling constant $C$ increases. Note that a soliton is
unstable if there is some $\Omega$ with either $\Omega<0$ or
Im$(\Omega)\neq0$. }\label{sketch1}

\end{figure}

We begin by considering bright soliton solutions, for which $u_n\to0$ as
$n\to\pm\infty$. This then implies that (for $C=0$) the eigenvalue
$\Omega^{(0)}_C$ has infinite multiplicity; it generates a corresponding
continuous spectrum (phonon band) for finite positive $C$. To
investigate the significance of this continuous spectrum, we introduce a
plane wave expansion
 \[
\eta_n=ae^{i\kappa n}+be^{-i\kappa n},
 \]
from which one obtains the dispersion relation
 \begin{equation}\label{dispb}
    \Omega=(2C(\cos\kappa-1)-\Lambda)^{2}-\gamma^{2}.
 \end{equation}
This in turn shows that the continuous band lies between
 \begin{equation}\label{omegabL}
\Omega_{L}=\Lambda^{2}-\gamma^{2}, \mbox{ when $\kappa=0$},
 \end{equation}
and
 \begin{equation}\label{omegabU}
\Omega_{U}=\Lambda^{2}-\gamma^{2}+8C(\Lambda+2C),
\mbox{ when $\kappa=\pi$.}
 \end{equation}

From the continuous spectrum analysis above, it can be concluded that an
instability can only be caused by the dynamics of discrete spectrum.


\subsubsection{Onsite bright solitons}

The existence and stability of a single excited state, i.e.\ an onsite
bright soliton, in the presence of a parametric driving has been
considered in Ref.~\onlinecite{susa06}. For small $C$, the soliton is
given by \cite{susa06}
 \begin{equation}
u_n=\left\{\begin{array}{ll}
\sqrt{\Lambda+\gamma}+C/\sqrt{\Lambda+\gamma}+\Ord(C^2),&n=0,\\
C/\sqrt{\Lambda+\gamma}+\Ord(C^2),&n=-1,1,\\
\Ord(C^2),&\mbox{otherwise},
\end{array}
\right.
\label{ds1}
 \end{equation}
and its eigenvalue by
 \begin{equation}
\Omega_E=4(\Lambda+\gamma)\gamma+8\gamma C+\Ord(C^2).
 \end{equation}

It was shown in Ref.~\cite{susa06} that the configuration \eqref{ds1},
which is known to be stable for any value of $C$ when $\gamma=0$, can be
destabilized by parametric driving. Furthermore, it was shown that there
are two mechanisms of destabilization, as sketched in
Fig.~\ref{sketch1}. The two instability scenarios are determined by the
relative positions of $\Omega_E^{(0)}$ and $\Omega_C^{(0)}$, as we now
summarize. First, we note that there is a threshold value,
$\gamma_{\rm{th}}=\Lambda/5$, at which the two leading-order eigenvalues
coincide, so that $\Omega_E^{(0)}=\Omega_C^{(0)}$. For
$\gamma>\gamma_{\rm{th}}$, upon increasing $C$ from $C=0$, the
instability is caused by the collision of $\Omega_E$ with $\Omega_U$;
taking $\Omega_E=\Omega_U$ then yields the corresponding approximate
critical value
 \begin{eqnarray}
\gamma^1_{\text{cr}}&=&
-\frac25\Lambda-\frac45 C+\frac15\sqrt{9\Lambda^2+56C\Lambda+96C^2}.
\label{gc2}
 \end{eqnarray}
For $\gamma<\gamma_{\rm{th}}$, by contrast, the instability is caused by
the collision of $\Omega_E$ with an eigenvalue bifurcating from
$\Omega_L$. In this case, the critical value of $\gamma$ can be
approximated by taking $\Omega_E=\Omega_L$, giving
 \begin{eqnarray}
\gamma^2_{\text{cr}}&=&
-\frac25\Lambda-\frac45 C+\frac15\sqrt{9\Lambda^2+16C(\Lambda+C)}.
\label{gc1}
 \end{eqnarray}
Together, $\gamma^1_{\text{cr}}$ and $\gamma^2_{\text{cr}}$ give
approximate boundaries of the instability region in the
$(C,\gamma)$-plane.

\begin{figure}[tbhp]
\centering
\includegraphics[width=6cm,clip=]{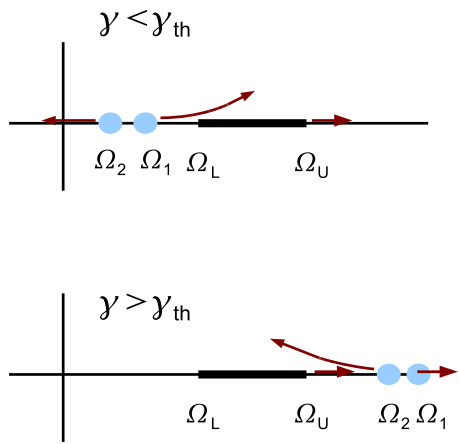}

\caption{As Fig.~\ref{sketch1}, but for a stable intersite bright
soliton.}\label{sketch2}

\end{figure}

\subsubsection{Intersite bright solitons}

The next natural fundamental solution to be considered is an intersite
bright soliton, i.e., a two-excited-site discrete mode. In the uncoupled
limit, the mode structure is of the form $u_{n}^{(0)}=0$ for $n\neq0,1$
and $u_{0}^{(0)}=u_{1}^{(0)}=\sqrt{\Lambda+\gamma}$. Using a
perturbative expansion, one can show further that the soliton is given
by
 \begin{equation}\label{intersitebright}
    u_{n}=\left\{
    \begin{array}{ll}
    \sqrt{\Lambda+\gamma}+\frac12C/\sqrt{\Lambda+\gamma}+\Ord(C^2), &
    n=0,1, \\
    C/\sqrt{\Lambda+\gamma}+\Ord(C^2), & n=-1,2, \\
    \Ord(C^2), & \mbox{otherwise.} \\
                   \end{array}
                 \right.
 \end{equation}

To study the stability of the intersite bright soliton above, let us
consider the ${\Ord}(1)$ equation \eqref{o0}. Due to the presence of
two non-zero excited sites at $C=0$, the soliton \eqref{intersitebright}
has at leading order the double eigenvalue
$\Omega^{(0)}_E=4(\Lambda+\gamma)\gamma$, with corresponding
eigenvectors $\eta^{(0)}_n=0$ for $n\neq0,1$, $\eta^{(0)}_0=1$, and
$\eta^{(0)}_1=\pm1$.

The continuation of the eigenvalue $\Omega^{(0)}_E$ above for nonzero
coupling $C$ can be obtained from Eq.~\eqref{o1} by applying a
solvability condition. The Fredholm alternative requires that $f_n=0$
for all $n$, from which we immediately deduce that the double eigenvalue
splits into two distinct eigenvalues, which are given as functions of
$C$ by
 \begin{equation}\label{eigbrightd1}
    \Omega_{1}=4(\Lambda+\gamma)\gamma+4\gamma C+\Ord(C^2),
 \end{equation}
and
 \begin{equation}\label{eigbrightd2}
    \Omega_{2}=4(\Lambda+\gamma)\gamma-4(\Lambda+\gamma)C+\Ord(C^2).
 \end{equation}

As is the case for onsite discrete solitons, intersite bright solitons
can also become unstable. The mechanism of the instability is again
determined by the relative positions of $\Omega_E^{(0)}$ and
$\Omega^{(0)}_C$, as sketched in Fig.~\ref{sketch2}. Performing an
analysis corresponding to that in Ref.~\cite{susa06}, we find that the
two mechanisms of destabilization for an onsite discrete soliton also
occur here. The two scenarios have corresponding critical values of
$\gamma$, which are given as functions of $C$ by
  \begin{eqnarray}
    \gamma_{{\rm cr}}^1=
-\frac{2}{5}\Lambda + \frac{2}{5}C +
\frac{1}{5}\sqrt{9\Lambda^{2}+52\Lambda C+84C^{2}}, \label{gammaP1a}\\
    \gamma_{{\rm cr}}^{2}=
-\frac{2}{5}\Lambda - \frac{2}{5}C+
\frac{1}{5}\sqrt{9\Lambda^{2}+8\Lambda C+4C^{2}}. \label{gammaP1b}
  \end{eqnarray}

We emphasize, as is apparent from the sketch shown in
Fig.~\ref{sketch2}, that there is another possible mechanism of
destabilization for $\gamma<\gamma_{{\rm th}}$, namely when $\Omega_2$
becomes negative. The third critical choice of parameter values is then
obtained by setting $\Omega_2=0$, i.e.
    \begin{equation}\label{gammaP2}
    \gamma_{{\rm cr}}^{3}=C.
  \end{equation}

\subsection{Comparisons with numerical calculations}

We have solved the steady-state equation \eqref{schr2} numerically using
a Newton--Raphson method, and analyzed the stability of the numerical
solution by solving the eigenvalue problem \eqref{lin2}. In this
section, we compare these numerical results with the analytical
calculations of the previous section. For the sake of simplicity, we set
$\Lambda=1$ in all the illustrative examples.

\subsubsection{Onsite bright solitons}

\begin{figure}[tbhp]
\centering
\includegraphics[width=8cm]{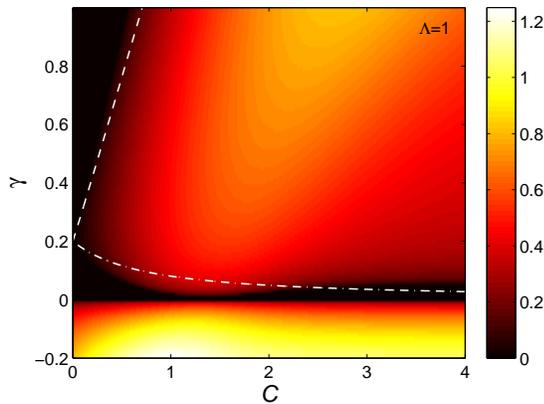}

\caption{(Colour online) The (in)stability region of onsite bright
solitons in $(C,\gamma)$ space. For each value of $C$ and $\gamma$, the
corresponding colour indicates the maximum value of $|$Im$(\omega)|$
(over all eigenvalues $\omega$) for the steady-state solution at that
point. Stability is therefore indicated by the region in which
Im$(\omega)=0$, namely the dark region. White dashed and dash-dotted
lines give the analytical approximations \eqref{gc2} and \eqref{gc1},
respectively.} \label{figbright6}

\end{figure}

Comparisons between numerical calculations and analytical approximations
for the case of onsite bright solitons have been fully presented and
discussed in Ref.~\cite{susa06}. For the sake of completeness, we
reproduce the results of Ref.~\cite{susa06} for the (in)stability domain
of onsite solitons in the $(C,\gamma)$ plane in Fig.~\ref{figbright6}.
Approximations \eqref{gc2} and \eqref{gc1} are also shown there.

\subsubsection{Intersite bright solitons}

\begin{figure}[tbhp]
\centering
\includegraphics[width=8cm]{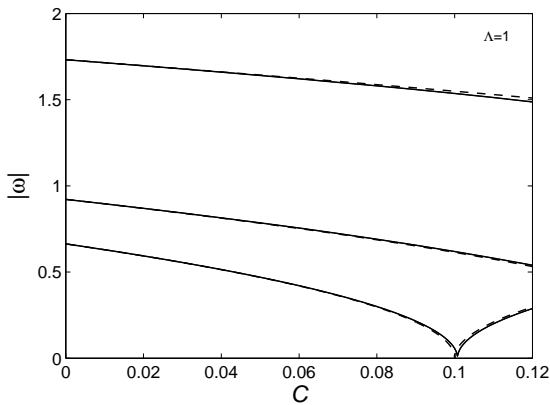}

\caption{Comparison between the critical eigenvalue of intersite bright
solitons obtained numerically (solid lines) and its analytical
approximation (dashed lines). The upper and lower curves correspond,
respectively, to $\gamma=0.5$ and $\gamma=0.1$, approximated by
Eq.~(\ref{eigbrightd2}), whereas the middle one corresponds to
$\gamma=0.18$, approximated by
Eq.~(\ref{eigbrightd1}).}\label{figbright2}

\end{figure}

\begin{figure*}[tbhp]
\centering
\subfigure[$\gamma=0.1,\,C=0.05$]
{\includegraphics[width=8cm]{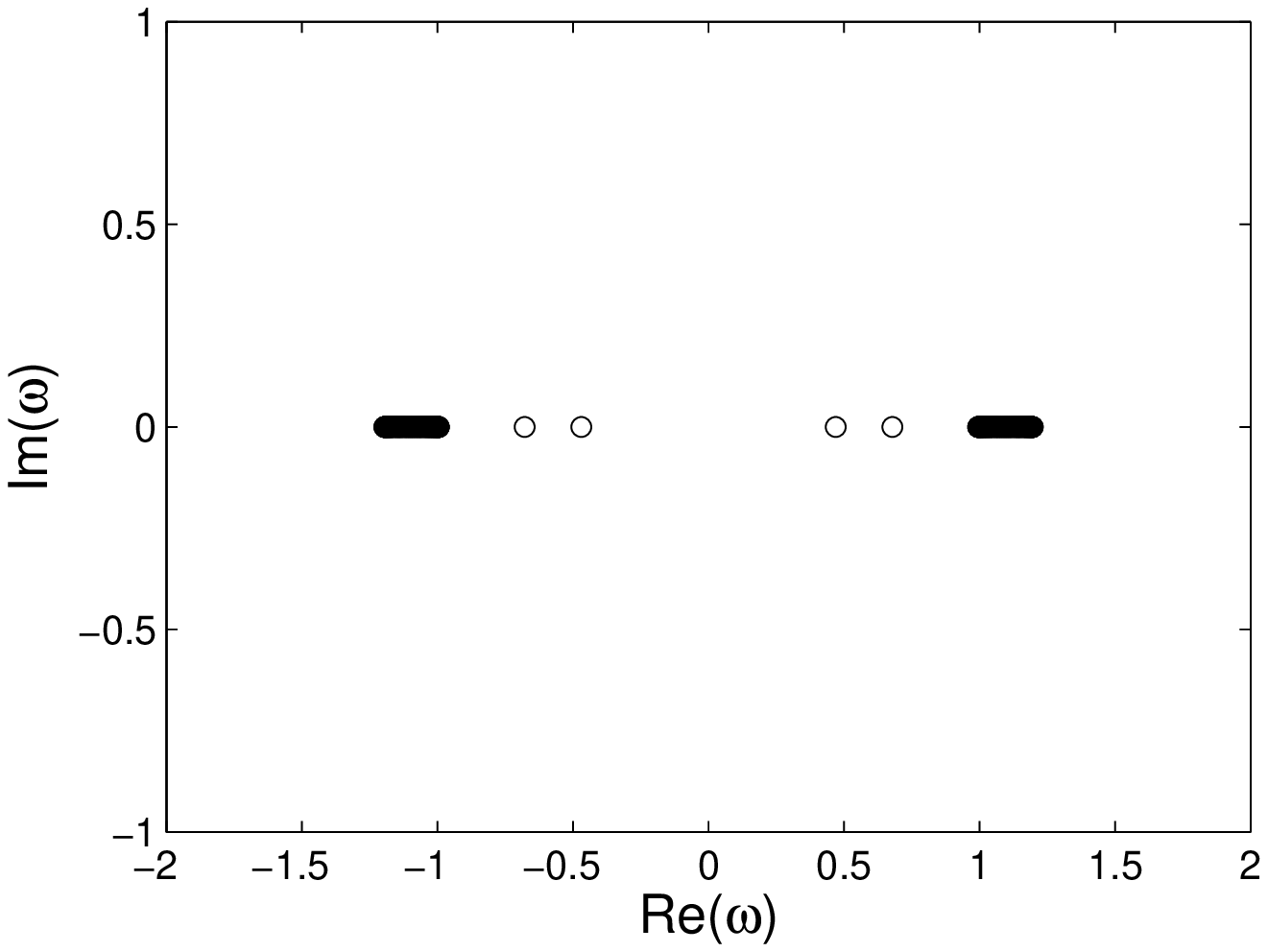}}
\subfigure[$\gamma=0.1,\,C=0.3$]
{\includegraphics[width=8cm]{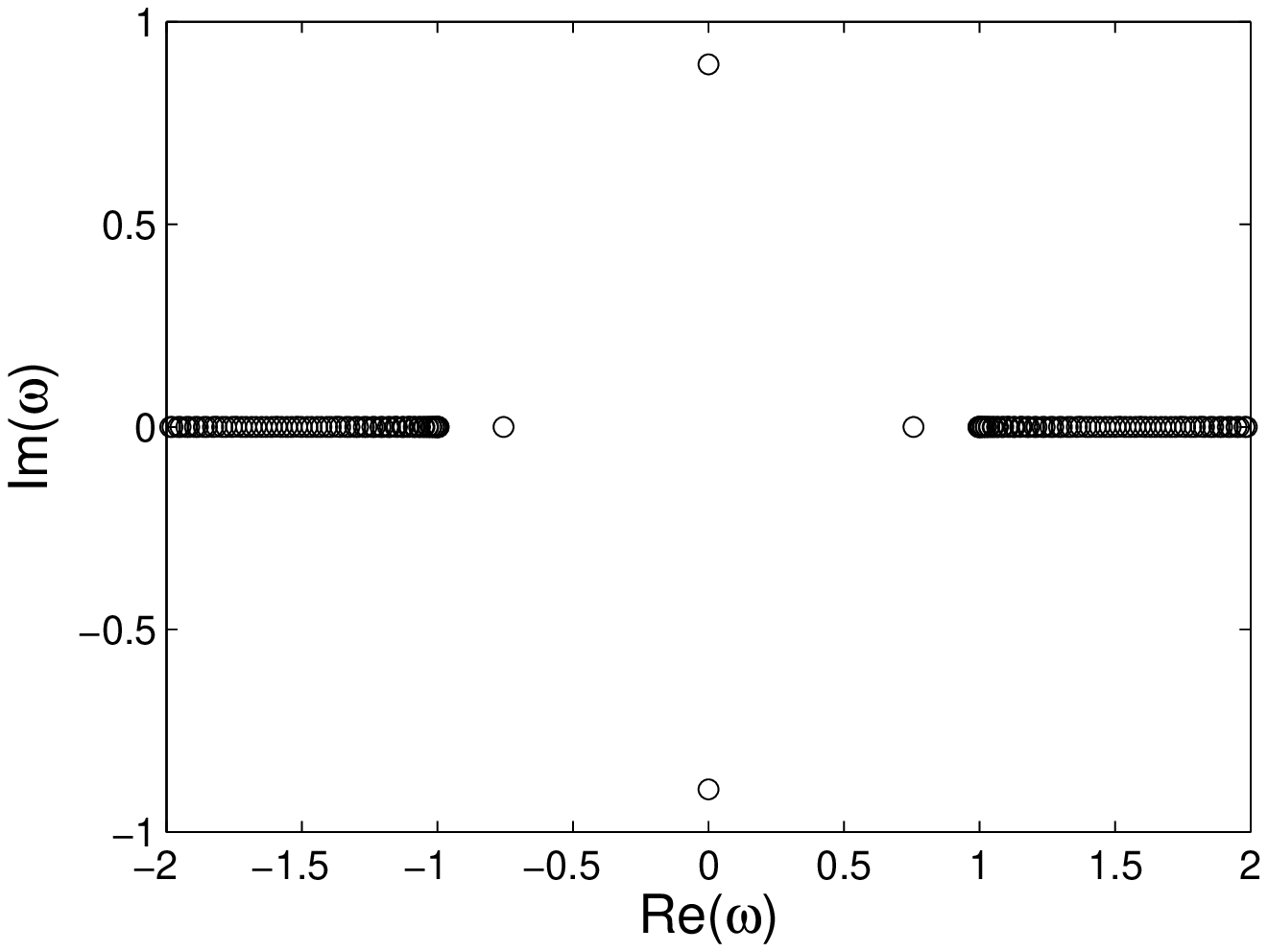}}\\
\subfigure[$\gamma=0.18,\,C=0.05$]
{\includegraphics[width=8cm]{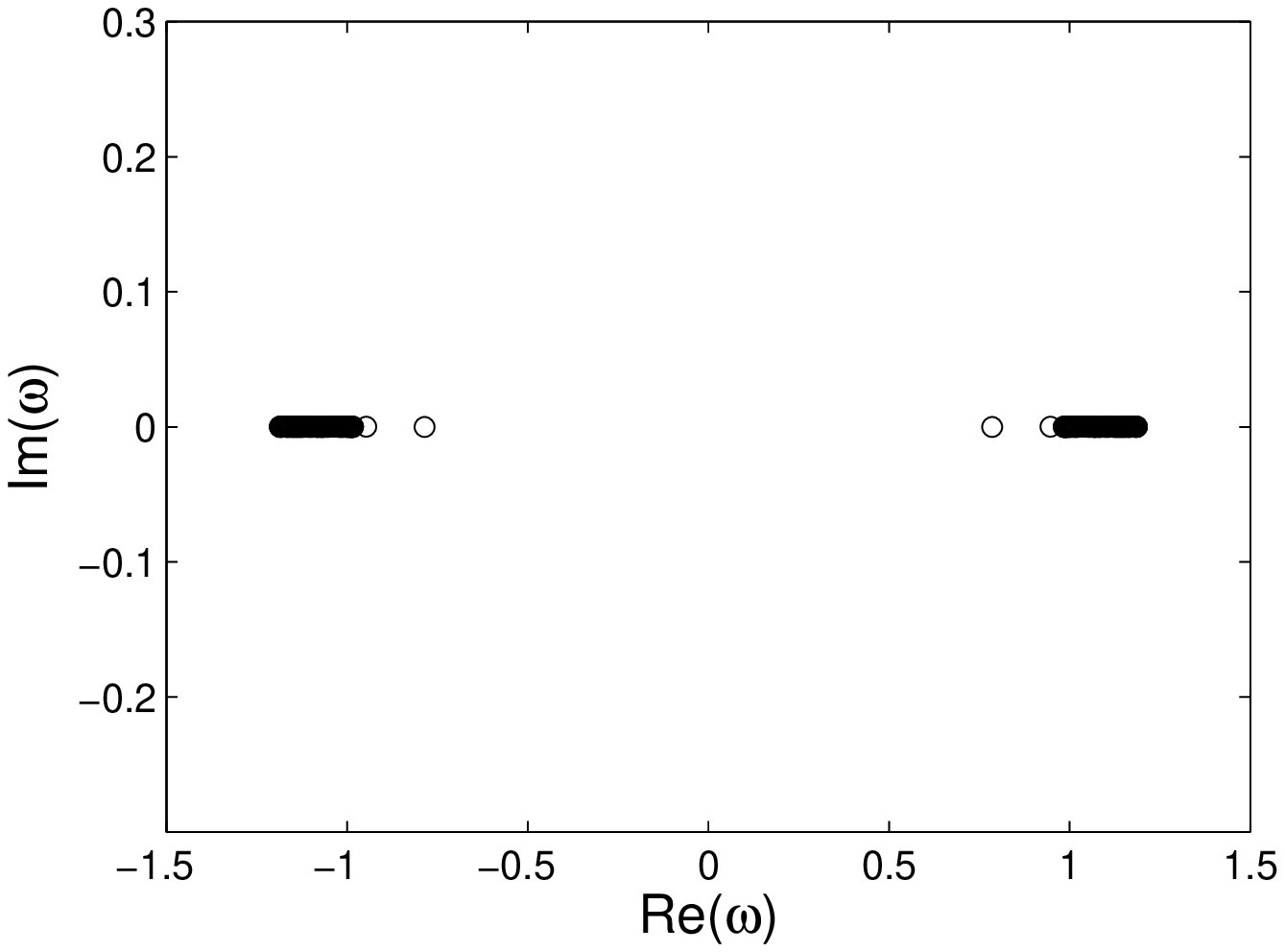}}
\subfigure[$\gamma=0.18,\,C=0.18$]
{\includegraphics[width=8cm]{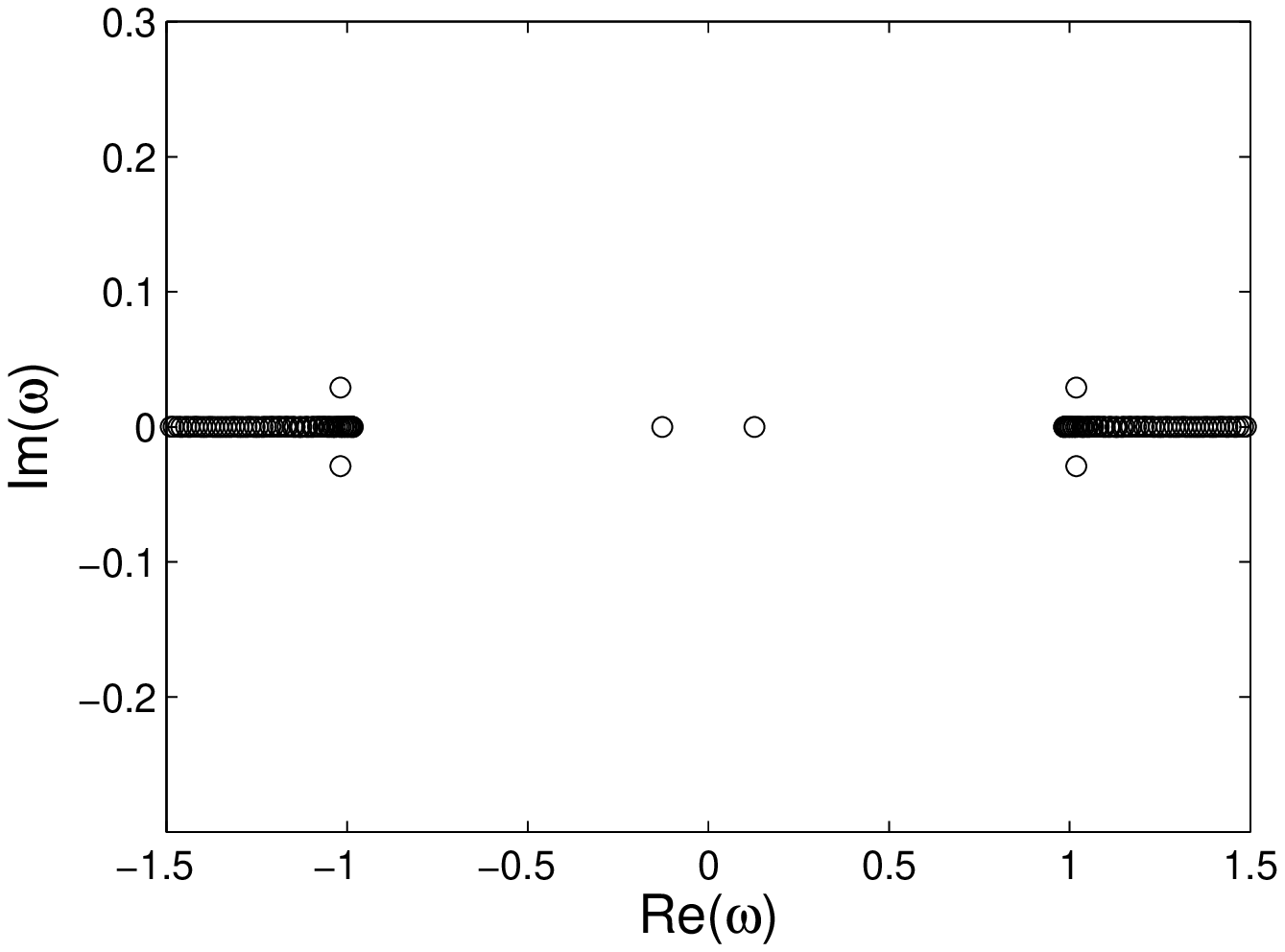}}\\
\subfigure[$\gamma=0.5,\,C=0.05$]
{\includegraphics[width=8cm]{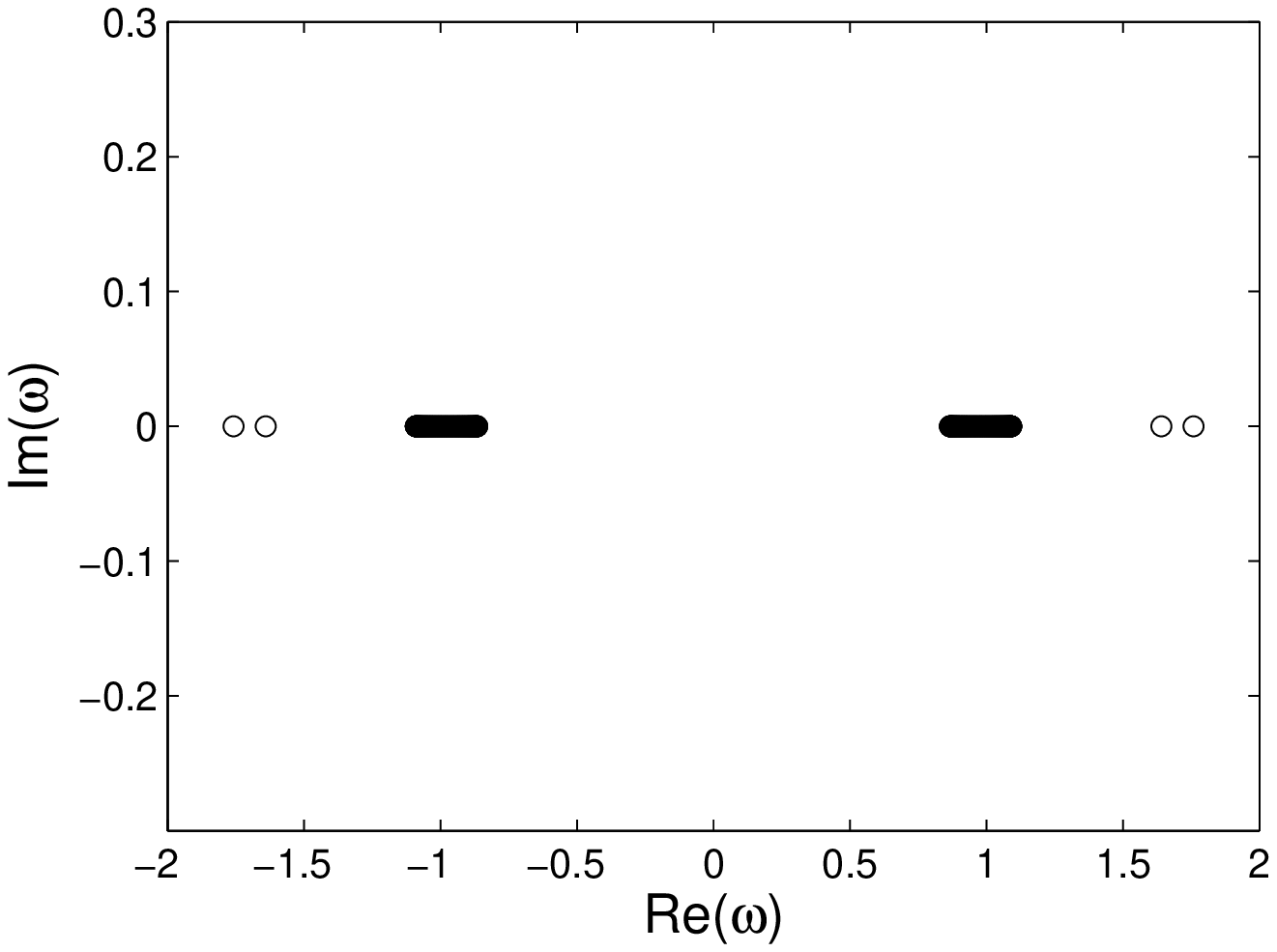}}
\subfigure[$\gamma=0.5,\,C=0.2$]
{\includegraphics[width=8cm]{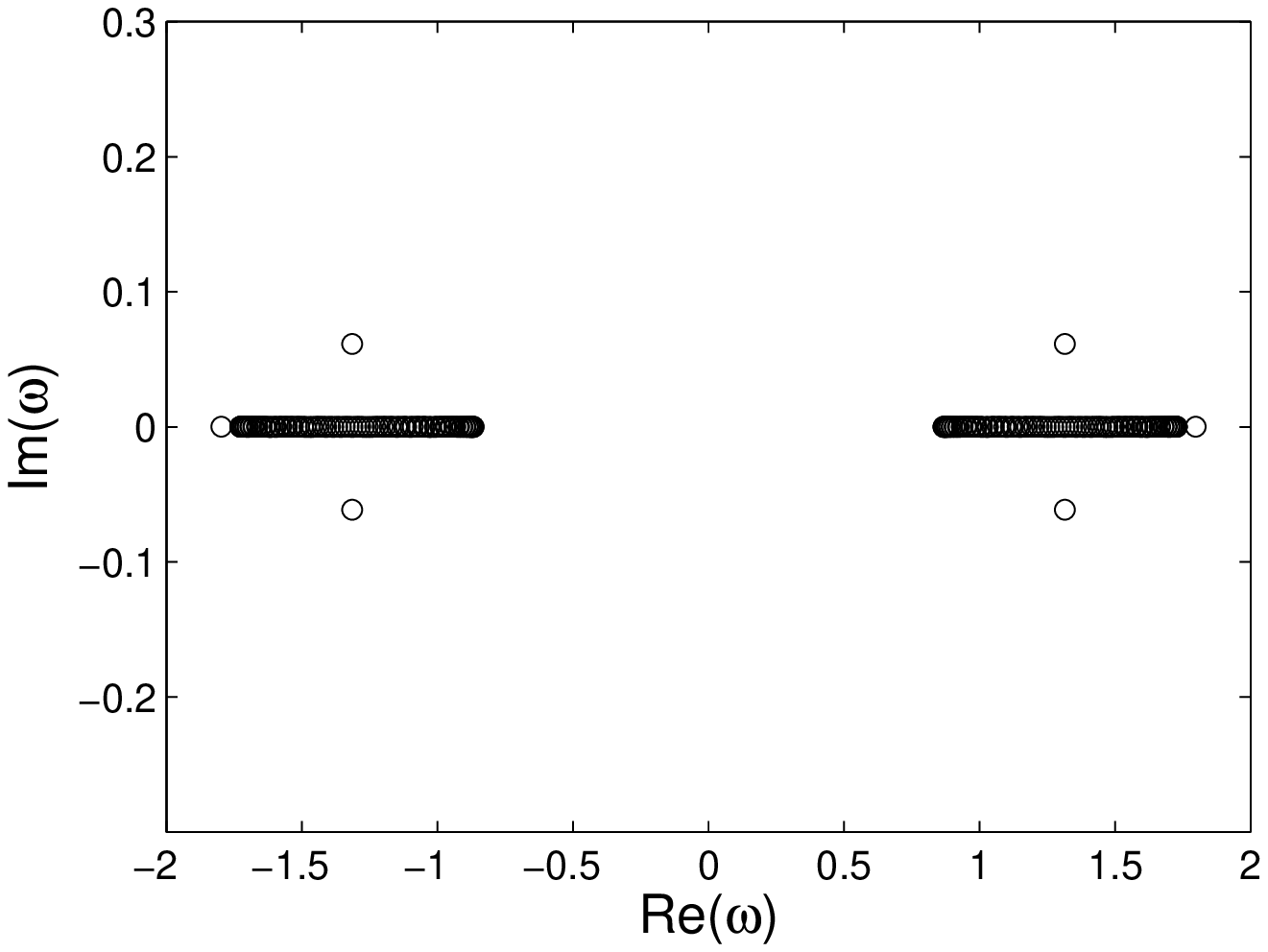}}

\caption{The structure of the eigenvalues of intersite bright solitons
in the complex plane for three values of $\gamma$, as indicated in the
caption of each plot. Left and right panels depict the eigenvalues of
stable and unstable solitons, respectively.}\label{figbright3}

\end{figure*}

\begin{figure}[tbhp]
\centering
\includegraphics[width=8cm]{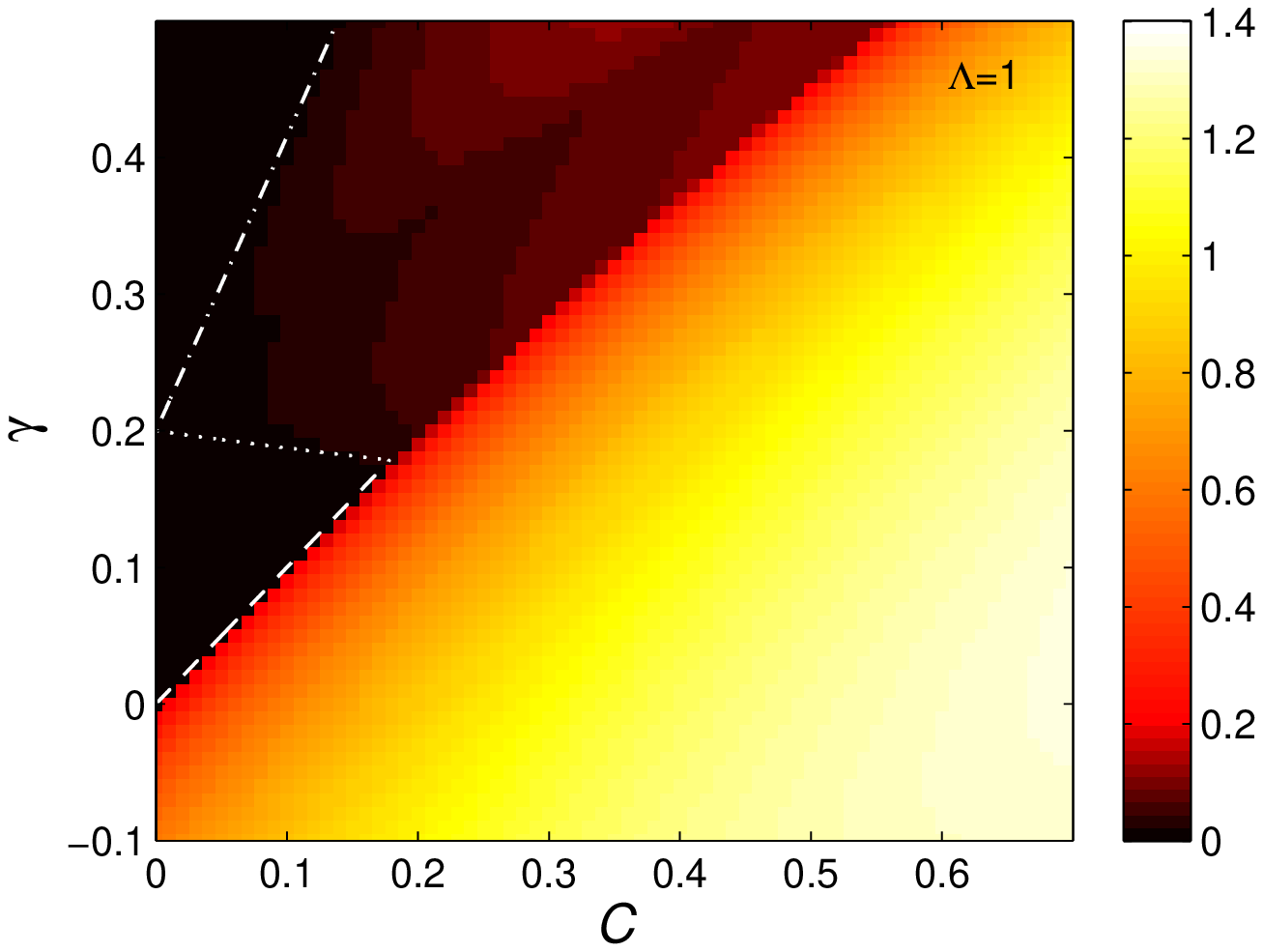}

\caption{(Colour online) As Fig.~\ref{figbright6}, but for intersite
bright solitons. Our analytical approximations, given by
Eqs.~(\ref{gammaP1a}), (\ref{gammaP1b}) and (\ref{gammaP2}), are shown
as white dash-dotted, dotted, and dashed lines, respectively.}
\label{figbright5}

\end{figure}

For the stability of intersite bright solitons, we start by examining
the validity of our analytical prediction for the eigenvalue associated
with the phase mode as given by Eqs.~(\ref{eigbrightd1}) and
(\ref{eigbrightd2}). In Fig.~\ref{figbright2}, we present a comparison
between the analytical approximation and the numerics for some
representative values of $\gamma$ (specifically $\gamma=0.1,0.18,0.5$).
This figure reveals the relative accuracy of the small-$C$
approximations, and we conclude that their range of validity is wider
for smaller values of~$\gamma$.

Next we turn to a description of the eigenvalue structure of this
intersite configuration for the three values of $\gamma$ given above;
this is shown in Fig.~\ref{figbright3}, where the left and right panels
respectively present the structure just before and just after the first
collision that results in the mode instability. We now describe results
in more detail for the three values of $\gamma$ in turn.

For $\gamma=0.1$, when $C=0$ the eigenvalues $\omega$ lie in the gap
between the two parts of the continuous spectrum, and the instability is
caused by a collision between the critical eigenvalue and its twin at
the origin (see the top panels of Fig.~\ref{figbright3}). For
$\gamma=0.18$, the eigenvalues $\omega$ also lie in the gap between the
two parts of the continuous spectrum, but the instability in this case
is due to a collision between one of the eigenvalues and the inner edge
of the continuous spectrum at $\omega=\pm\sqrt{\Omega_{L}}$ (see the
middle panels of Fig.~\ref{figbright3}). In contrast to the two cases
above, for $\gamma=0.5$ the eigenvalues lie beyond the continuous
spectrum, and the instability is caused by a collision between the
critical eigenvalue and the outer boundary at
$\omega=\pm\sqrt{\Omega_{U}}$ (see the bottom panels of
Fig.~\ref{figbright3}). All the numerical results presented here are in
accordance with the sketch shown in Fig.~\ref{sketch2}.

Numerical calculations of the stability of intersite bright solitons,
for a relatively large range of $C$ and $\gamma$, give us the stability
domain of the bright solitons in the two-parameter $(C,\gamma)$ plane,
which is presented in Fig.~\ref{figbright5}. We use colours to represent
the maximum of $|$Im$(\omega)|$ as a function of $C$ and $\gamma$; thus
solitons are stable in the black region. Our analytical predictions for
the occurrence of instability, given by
Eqs.~(\ref{gammaP1a})--(\ref{gammaP2}), are also shown, respectively, by
dashed, dotted, and dash-dotted lines.


\section{Dark solitons in the defocusing DNLS}

In this section we consider the existence and stability of onsite and
intersite dark solitons for the defocusing DNLS equation. Then a static
(real-valued, time-independent) solution $u_n$ of \eqref{gov} satisfies
 \begin{equation}
-C\Delta_2u_n+u_n^3-\Lambda u_n-\gamma u_n=0.
\label{schr3}
 \end{equation}
In contrast to bright solitons, where $u_n\to0$ as $n\to\pm\infty$, dark
solitons have $u_n\to\pm\sqrt{\Lambda+\gamma}$ as $n\to\pm\infty$.

To examine the stability of $u_n$, we again introduce the linearization
ansatz $\phi_{n}=u_{n}+\delta\epsilon_{n}$, where again $\delta\ll1$.
Substituting this ansatz into the defocusing equation (\ref{gov}),
writing $\epsilon_{n}(t)=\eta_{n}+i\xi_{n}$, and linearizing in $\delta$,
we  again find
 \begin{equation}\label{matrix}
    \left(
      \begin{array}{c}
        \dot{\eta}_{n} \\
        \dot{\xi}_{n} \\
      \end{array}
    \right)=\left(
              \begin{array}{cc}
                0 & \mathcal{L}_+ \\
                -\mathcal{L}_- & 0 \\
              \end{array}
            \right)\left(
                     \begin{array}{c}
                       \eta_{n} \\
                       \xi_{n} \\
                     \end{array}
                   \right)=\mathcal{H}\left(
                                        \begin{array}{c}
                                          \eta_{n} \\
                                          \xi_{n} \\
                                        \end{array}
                                      \right),
 \end{equation}
but where the operators $\mathcal{L}_{\pm}(C)$ are now defined as
 \begin{eqnarray*}
\mathcal{L}_{-}(C)&\equiv&-C\Delta_{2}+(3u_{n}^{2}-\Lambda-\gamma),\\
\mathcal{L}_{+}(C)&\equiv&-C\Delta_{2}+(u_{n}^{2}-\Lambda+\gamma).
 \end{eqnarray*}
The eigenvalue problem above can be simplified further as for the
focusing case, to the alternative form
 \begin{equation}\label{eigprob}
\mathcal{L}_{+}(C)\mathcal{L}_{-}(C)\eta_{n}=
\omega^{2}\eta_{n}=\Omega\eta_{n}.
 \end{equation}
Performing a stability analysis as before, we find the dispersion relation
for a dark soliton to be
 \begin{equation}\label{disp}
    \Omega=(2C(\cos\kappa-1)-(\Lambda+2\gamma))^{2}-\Lambda^{2},
 \end{equation}
and so the continuous band lies between
 \begin{equation}\label{omegaL}
\Omega_{L}=4(\Lambda+\gamma)\gamma, \mbox{ when $\kappa=0$},
 \end{equation}
and
 \begin{equation}\label{omegaU}
\Omega_{U}=4(\Lambda+\gamma)\gamma+8C(\Lambda+2\gamma+2C),
\mbox{ when $\kappa=\pi$.}
 \end{equation}

\subsection{Analytical calculations}

To study the eigenvalue(s) of the dark soliton analytically, we again
expand $\eta_n$ and $\Omega$ in powers of $C$, and hence obtain from
(\ref{eigprob}), at ${\Ord}(C^0)$ and ${\Ord}(C^1)$,
respectively, the equations
 \begin{equation}\label{Lo}
\left[\mathcal{L}_{+}(0)\mathcal{L}_{-}(0)-\Omega^{(0)}\right]
\eta_{n}^{(0)}=0,
\end{equation}
and
\begin{equation}\label{Lf}
\left[\mathcal{L}_{+}(0)\mathcal{L}_{-}(0)-\Omega^{(0)}\right]
\eta_{n}^{(1)}=f_{n},
\end{equation}
with
\begin{equation}\label{fn}
f_{n}=(Q_{n}+\Omega^{(1)})\eta_{n}^{(0)},
\end{equation}
where
\begin{equation}\label{Qn}
Q_{n}=(\Delta_{2}-2u_{n}^{(0)}u_{n}^{(1)})\mathcal{L}_{-}(0)+
\mathcal{L}_{+}(0)(\Delta_{2}-6u_{n}^{(0)}u_{n}^{(1)}).
 \end{equation}
We next investigate the eigenvalues of both intersite and onsite modes.

\subsubsection{Onsite dark solitons}

With errors of order  $C^2$, an onsite dark soliton is given by
\begin{equation}\label{onsite}
    u_{n}=\left\{
                   \begin{array}{ll}
                     -\sqrt{\Lambda+\gamma}, & \mbox{$n=-2,-3,\ldots$,} \\
-\sqrt{\Lambda+\gamma}+\frac12C/\sqrt{\Lambda+\gamma}, & \mbox{$n=-1$,} \\
                     0, & \hbox{$n=0$,}\\
\sqrt{\Lambda+\gamma}-\frac12C/\sqrt{\Lambda+\gamma}, & \mbox{$n=1$,} \\
                     \sqrt{\Lambda+\gamma}, & \mbox{$n=2,3,\ldots$.} \\
                   \end{array}
                 \right.
\end{equation}
For this configuration,
\begin{equation}\label{LL}
\mathcal{L}_{+}(0)\mathcal{L}_{-}(0)=
                    \left\{
                       \begin{array}{ll}
                         \Lambda^{2}-\gamma^{2}, & \mbox{$n=0$,} \\
                        4(\Lambda+\gamma)\gamma, & \mbox{$n \neq 0$.}
                       \end{array}
                     \right. \end{equation}
From Eq.~(\ref{Lo}), we then deduce that at $C=0$ the eigenvalues of
onsite discrete dark solitons are given by
$\Omega^{(0)}_C=4(\Lambda+\gamma)\gamma$, which becomes the continuous
band for nonzero $C$, and $\Omega^{(0)}_E=\Lambda^{2}-\gamma^{2}$.

The continuation of the eigenvalue $\Omega^{(0)}_E$ for nonzero $C$ can
be calculated from Eq.\ (\ref{Lf}). The coefficient of $\eta_n^{(1)}$ in
this case is given by
 \begin{equation}\label{LLomegaon}
\mathcal{L}_{+}(0)\mathcal{L}_{-}(0)-\Omega^{(0)}=
                    \left\{
                       \begin{array}{ll}
                         0, & \mbox{$n=0$,} \\
               4\Lambda\gamma-\Lambda^{2}+5\gamma^{2}, & \mbox{$n \neq 0$.}
                       \end{array}
                     \right.
\end{equation}
The solvability condition for (\ref{Lf}) then requires that
$f_{0}=(4\Lambda-\Omega^{(1)})\eta_{0}^{(0)}=0$.  Setting
$\eta_{0}^{(0)}\neq0$, we deduce that $\Omega^{(1)}=-4\Lambda$.
Hence the eigenvalue of an onsite dark soliton for small $C$ is given by
 \begin{equation}\label{eigd}
    \Omega=\Lambda^{2}-\gamma^{2}-4\Lambda C+\Ord(C^2).
 \end{equation}

\begin{figure}[tbhp]
\centering
\includegraphics[width=6cm,clip=]{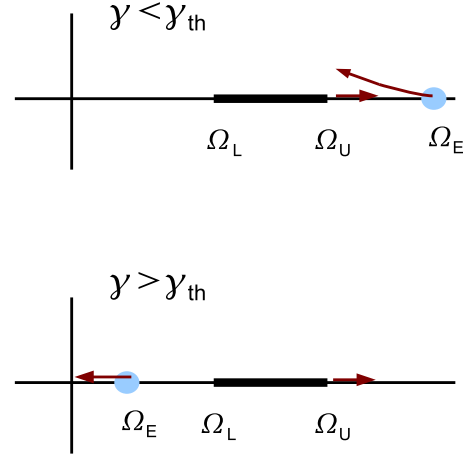}

\caption{As Fig.~\ref{sketch1}, but for a stable onsite dark
soliton.}\label{sketch3}

\end{figure}

Initially, i.e.\ for $C=0$, the relative positions of the eigenvalue and
the continuous spectrum can be divided into two cases, according to
whether $\gamma\gtrless\gamma_{{\rm th}}=\Lambda/5$. When $C=0$ and
$\gamma<\gamma_{{\rm th}}$ ($\gamma>\gamma_{{\rm th}}$) the eigenvalue
(\ref{eigd}) will be above (below) the continuous spectrum, as sketched
in Fig.~\ref{sketch3}. These relative positions determine the
instability mechanism for an onsite dark soliton, as we now describe.

For $\gamma<\Lambda/5$, the instability is due to a collision between
the eigenvalue (\ref{eigd}) and $\Omega_{U}$, which approximately occurs
when $\gamma=\gamma_{{\rm cr}}^{1}$, where
  \begin{equation}\label{gamma1}
    \gamma_{{\rm cr}}^{1}=
-\frac{2}{5}\Lambda - \frac{8}{5}C+
\frac{1}{5}\sqrt{9\Lambda^{2}-28\Lambda C-16C^{2}};
  \end{equation}
note that this critical value is meaningful only when
$C\leq9\Lambda/(14+2\sqrt{85})$. For $\gamma>\Lambda/5$, the instability
is caused by the eigenvalue (\ref{eigd}) becoming negative, which occurs
when $\gamma=\gamma_{{\rm cr}}^{2}$, where
  \begin{equation}\label{gamma2}
    \gamma_{{\rm cr}}^{2}=\sqrt{\Lambda^{2}-4\Lambda C};
  \end{equation}
this value is meaningful only when $C\leq\Lambda/4$.

Furthermore, if we include terms up to $\Ord(C^2)$, we obtain
\begin{equation}\label{eigdfurther}
    \Omega=\Lambda^{2}-\gamma^{2}-4\Lambda C+4C^{2}+\Ord(C^3)
\end{equation}
as the eigenvalue of an onsite discrete dark soliton. Using this expression,
we find the critical value of $\gamma$ indicating the onset of instability
to be
\begin{equation}\label{gamma1further}
    \gamma_{{\rm cr}}^{1}=-\frac{2}{5}\Lambda - \frac{8}{5}C+
\frac{1}{5}\sqrt{9\Lambda^{2}-28\Lambda C+4C^{2}},
\end{equation}
for $\gamma<0.2\Lambda$ and
\begin{equation}\label{gamma2further}
    \gamma_{{\rm cr}}^{2}=\sqrt{\Lambda^{2}-4\Lambda C+4C^{2}},
\end{equation}
for $\gamma\geq0.2\Lambda$.

\subsubsection{Intersite modes}

Intersite discrete dark solitons are given, with errors of ${\Ord}(C^2)$, by
 \begin{equation}\label{intersite}
    u_{n}=\left\{
                   \begin{array}{ll}
       -\sqrt{\Lambda+\gamma}, & \mbox{$n=-2,-3,\ldots$,} \\
       -\sqrt{\Lambda+\gamma}+C/\sqrt{\Lambda+\gamma}, & \mbox{$n=-1$,} \\
        \sqrt{\Lambda+\gamma}-C/\sqrt{\Lambda+\gamma}, & \mbox{$n=0$,} \\
                     \sqrt{\Lambda+\gamma}, & \mbox{$n=1,2,\ldots$}. \\
                   \end{array}
                 \right.
 \end{equation}

Starting from Eq.~(\ref{Lo}), we then find
 \begin{equation}
\mathcal{L}_{+}(0)\mathcal{L}_{-}(0)=4(\Lambda+\gamma)\gamma
 \end{equation}
for all $n$, from which we deduce that there is a single leading-order
eigenvalue, given by $\Omega^{(0)}=4(\Lambda+\gamma)\gamma$, with
infinite multiplicity. This eigenvalue then expands to form the continuous
spectrum for nonzero $C$.

Because a localized structure must have an eigenvalue, we infer that an
eigenvalue will bifurcate from the lower edge of the continuous
spectrum. This bifurcating eigenvalue may be calculated from
Eq.~(\ref{Lf}). Because
 \begin{equation}\label{LLomegaint}
\mathcal{L}_{+}^{(0)}(0)\mathcal{L}_{-}^{(0)}(0)-\Omega^{(0)}=0
 \end{equation}
for all $n$, the solvability condition for Eq.~(\ref{Lf}) requires
$f_{n}=0$ for all $n$. A simple calculation then yields
 \begin{equation}
f_{n}=\left\{
\begin{array}{lll}
\left[4\Lambda+16\gamma+(2\Lambda+4\gamma)\Delta_{2}+
\Omega^{(1)}\right]\eta_{n}^{(0)},&n=-1,0,\\
\left[(2\Lambda+4\gamma)\Delta_{2}+
\Omega^{(1)}\right]\eta_{n}^{(0)},&n\neq-1,0.
\end{array}
\right.
 \end{equation}
Taking $\eta_{n}^{(0)}=0$ for $n\neq-1,0$ leaves the two nontrivial
equations
 \begin{eqnarray*}
(8\gamma+\Omega^{(1)})\eta_{-1}^{(0)}+(2\Lambda+4\gamma)\eta_{0}^{(0)}&=&0,\\
(8\gamma+\Omega^{(1)})\eta_{0}^{(0)}+(2\Lambda+4\gamma)\eta_{-1}^{(0)}&=&0,
 \end{eqnarray*}
from which we see that $\eta_{-1}^{(0)}=\pm\eta_{0}^{(0)}$.
Thus we obtain two possibilities for the $\Ord(C)$ contribution to the
eigenvalue, given by
 \[
\Omega_{1}^{(1)}=-(12\gamma+2\Lambda),\quad
\Omega_{2}^{(1)}=2\Lambda-4\gamma.
 \]
Hence the eigenvalues bifurcating from the lower edge of the continuous
spectrum are given by
 \begin{equation}\label{eigdn}
    \Omega_{1}=4(\Lambda+\gamma)\gamma-(12\gamma+2\Lambda)C+\Ord(C^2),
 \end{equation}
and
 \begin{equation}\label{eigdp}
    \Omega_{2}=4(\Lambda+\gamma)\gamma+(2\Lambda-4\gamma)C+\Ord(C^2).
 \end{equation}

\begin{figure}[tbhp]
\centering
\includegraphics[width=6cm,clip=]{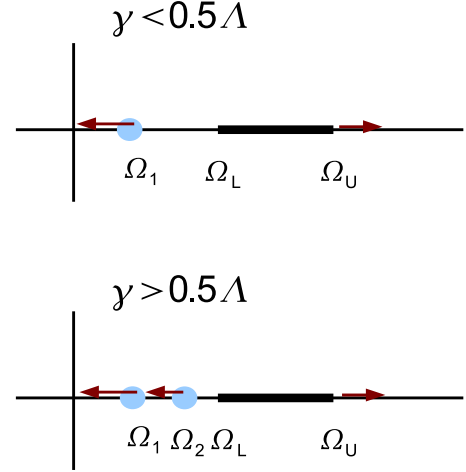}

\caption{As Fig.\ \ref{sketch1}, but for a stable intersite dark
soliton.}\label{sketch4}

\end{figure}

A simple analysis shows that $\Omega_2<\Omega_L$ only when
$\gamma>\Lambda/2$. The sketch in Fig.~\ref{sketch4} then illustrates
that instability is caused by $\Omega_1$ becoming negative. This
consideration gives the critical $\gamma$ as a function of the coupling
constant $C$ to be
 \begin{equation}\label{gammacr}
  \gamma_{{\rm cr}}=
-\frac{1}{2}\Lambda+\frac{3}{2}C+
\frac{1}{2}\sqrt{\Lambda^{2}-4\Lambda C+9C^{2}}.
 \end{equation}
When there are two eigenvalues ($\Omega_{1}$ and $\Omega_{2}$),
$\Omega_{2}$ decreases more slowly than $\Omega_{1}$, in such a way that
for $\gamma>\Lambda/2$ the instability is still caused by $\Omega_1$
becoming negative.

\subsection{Comparison with numerical computations}

\subsubsection{Onsite dark solitons}

\begin{figure}[tbhp]
\centering
\includegraphics[width=8cm]{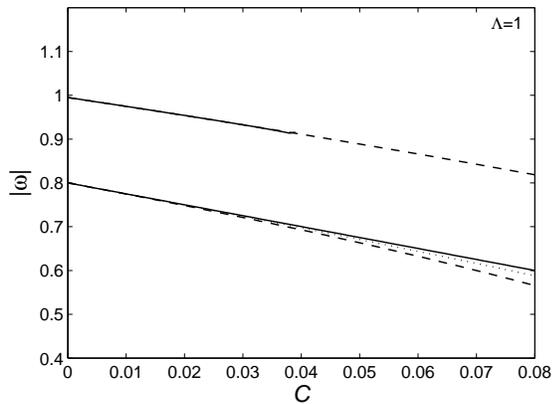}

\caption{Comparisons between the critical eigenvalue for on-site dark
solitons obtained numerically (solid lines) and analytically using Eq.\
(\ref{eigd}) (dashed lines) for $\gamma=0.1$ (upper curves) and
$\gamma=0.6$ (lower curves). An approximation that explicitly includes
the next term in expansion Eq.\ (\ref{eigdfurther}) is also plotted
(dotted lines).}\label{fig4}

\end{figure}

\begin{figure*}[tbhp] \centering
\subfigure[$\gamma=0.1,\,C=0.02$]
{\includegraphics[width=8cm]{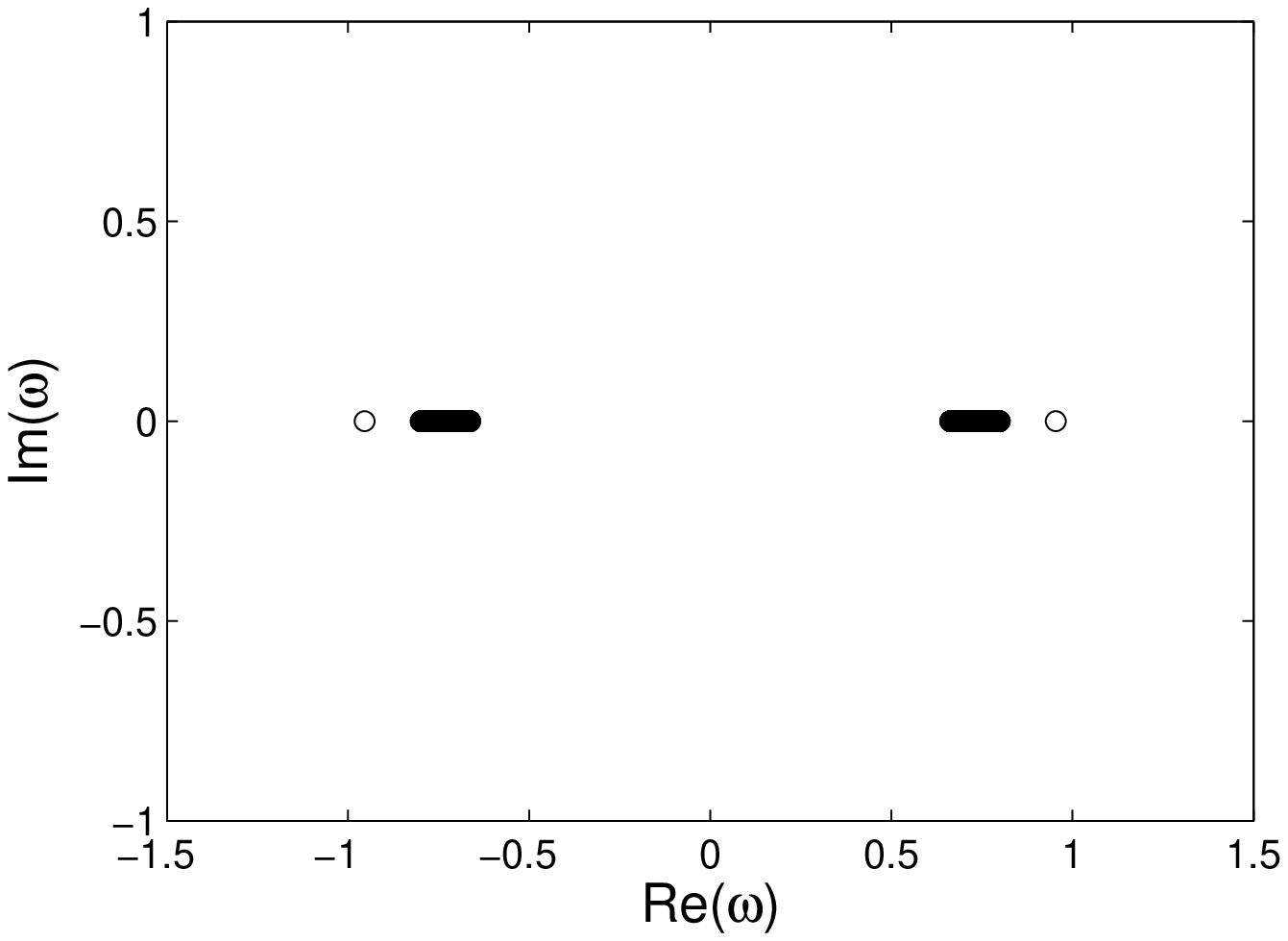}}
\subfigure[$\gamma=0.1,\,C=0.2$]
{\includegraphics[width=8cm]{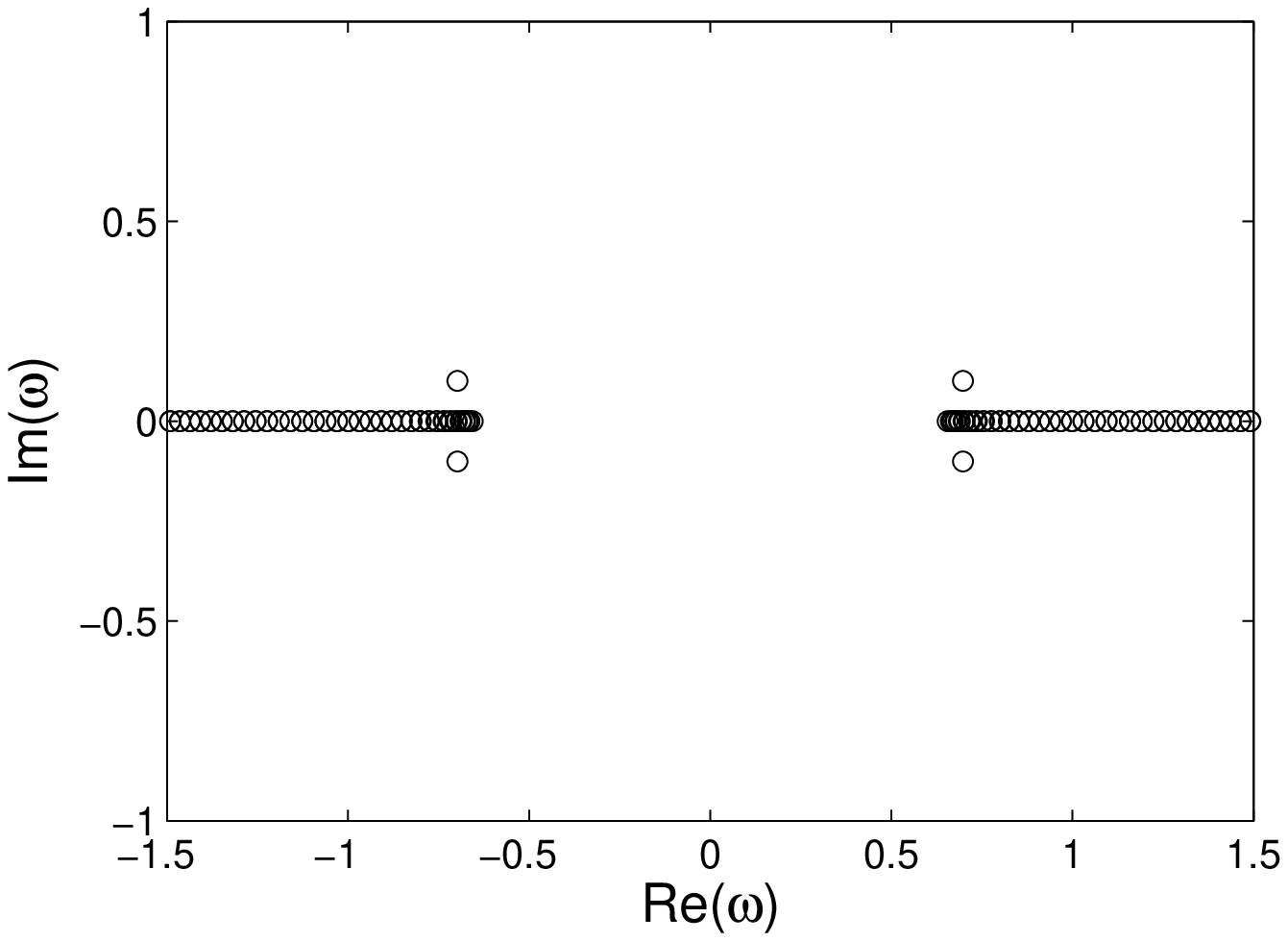}}\\
\subfigure[$\gamma=0.6,\,C=0.01$]
{\includegraphics[width=8cm]{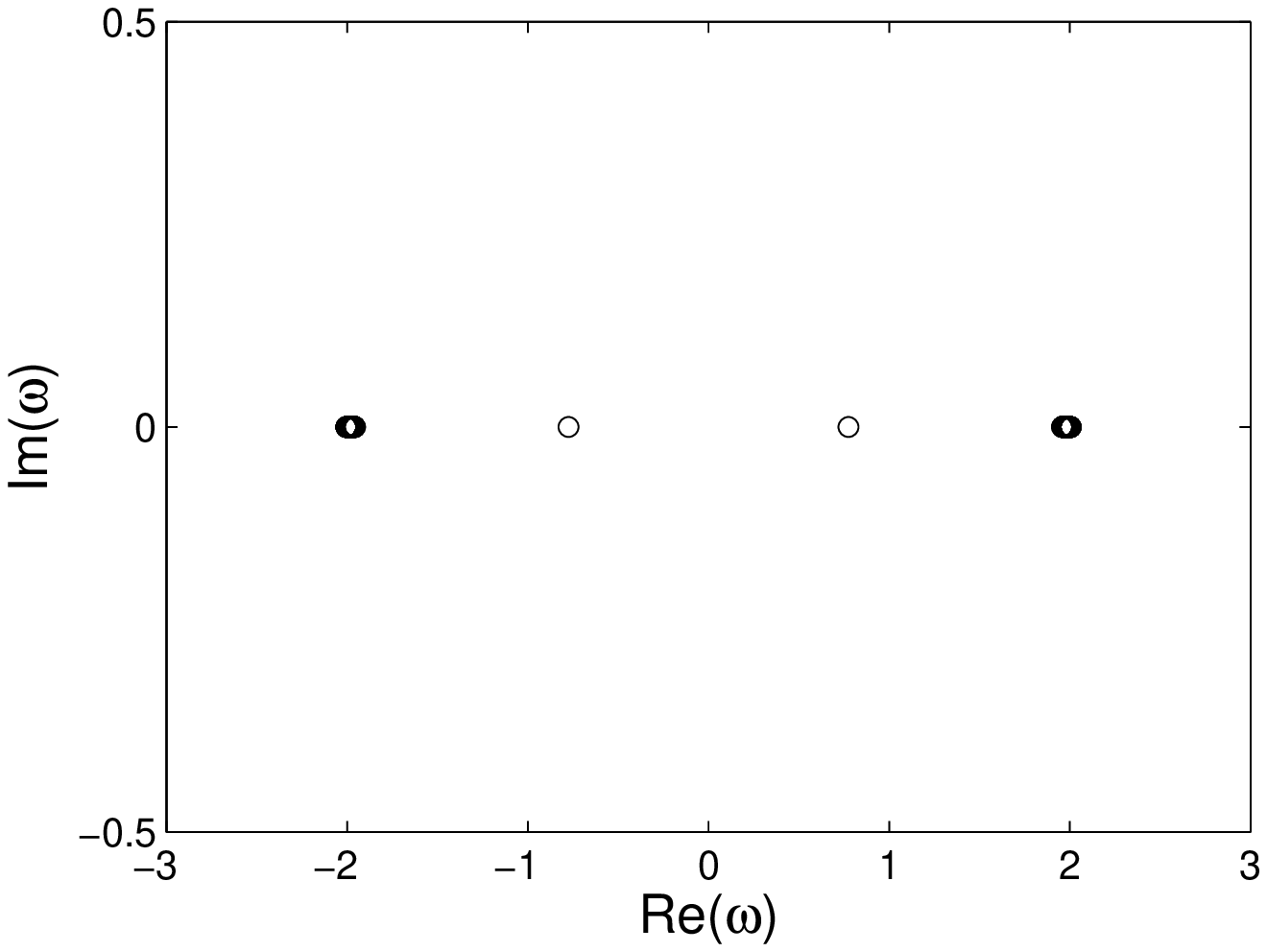}}
\subfigure[$\gamma=0.6,\,C=1$]
{\includegraphics[width=8cm]{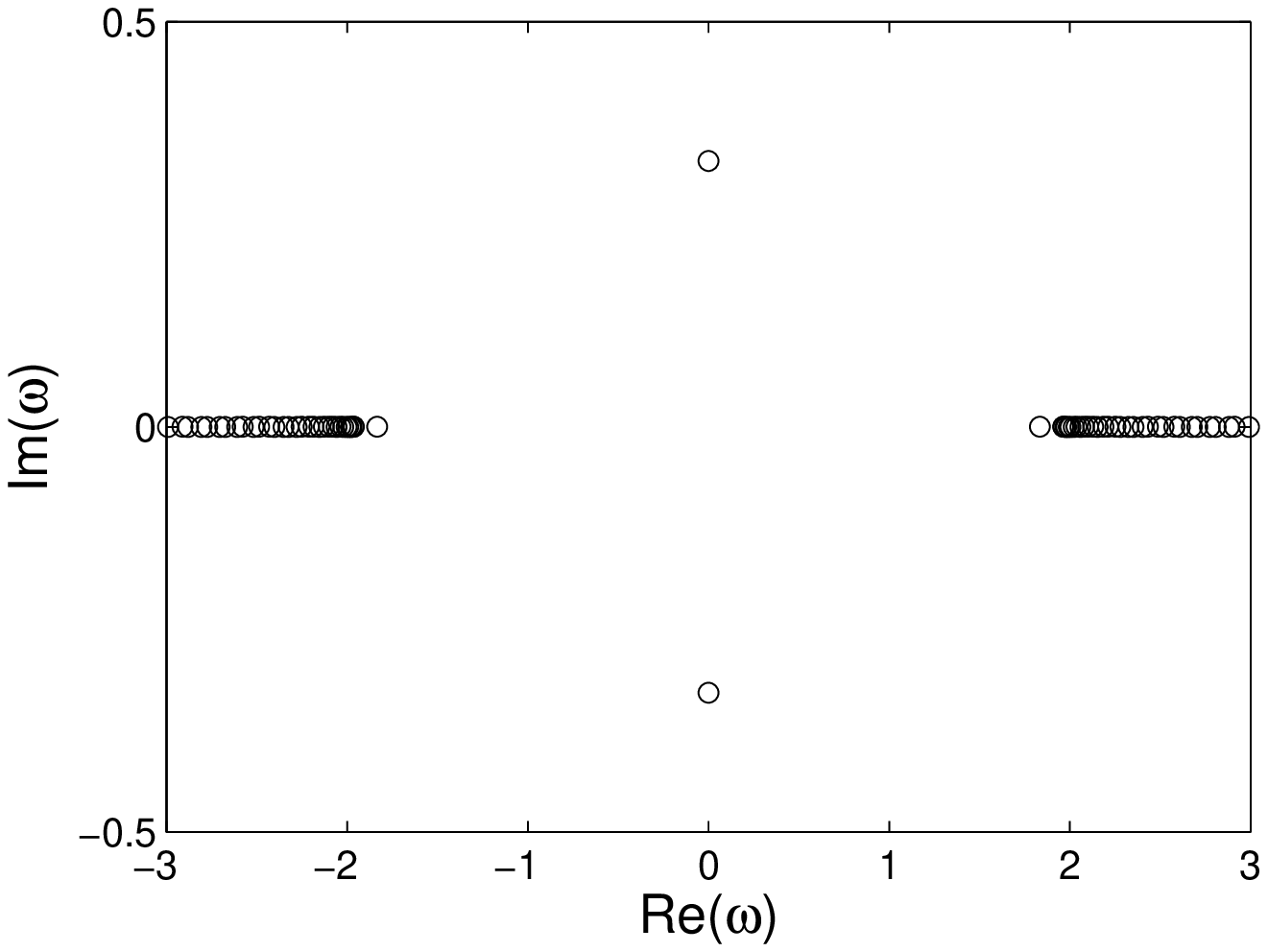}}

\caption{The eigenvalue structure of on-site dark solitons for several
values of $\gamma$ and $C$, as indicated in the caption of each
panel.}\label{fig5}

\end{figure*}

\begin{figure}[tbhp]
\centering
\includegraphics[width=8cm]{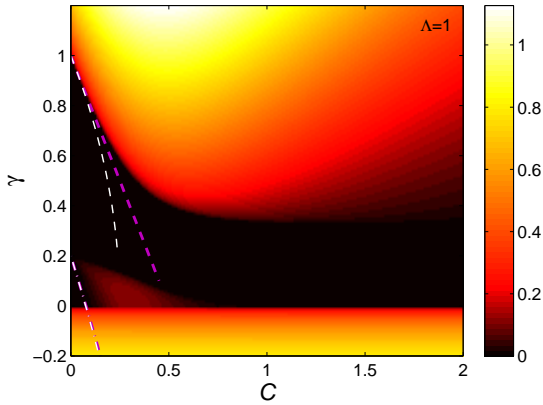}

\caption{(Colour online) The (in)stability region of onsite dark
solitons in the two-parameter $(C,\gamma)$ space. The white and pink
dashed lines respectively give the analytical approximations Eq.\
(\ref{gamma2}) and (\ref{gamma2further}). White and pink dash-dotted
lines show Eqs.~(\ref{gamma1}) and (\ref{gamma1further}); note that
these curves are indistinguishable in this plot.} \label{fig7}

\end{figure}

\begin{figure}[tbhp]
\centering
\includegraphics[width=8cm]{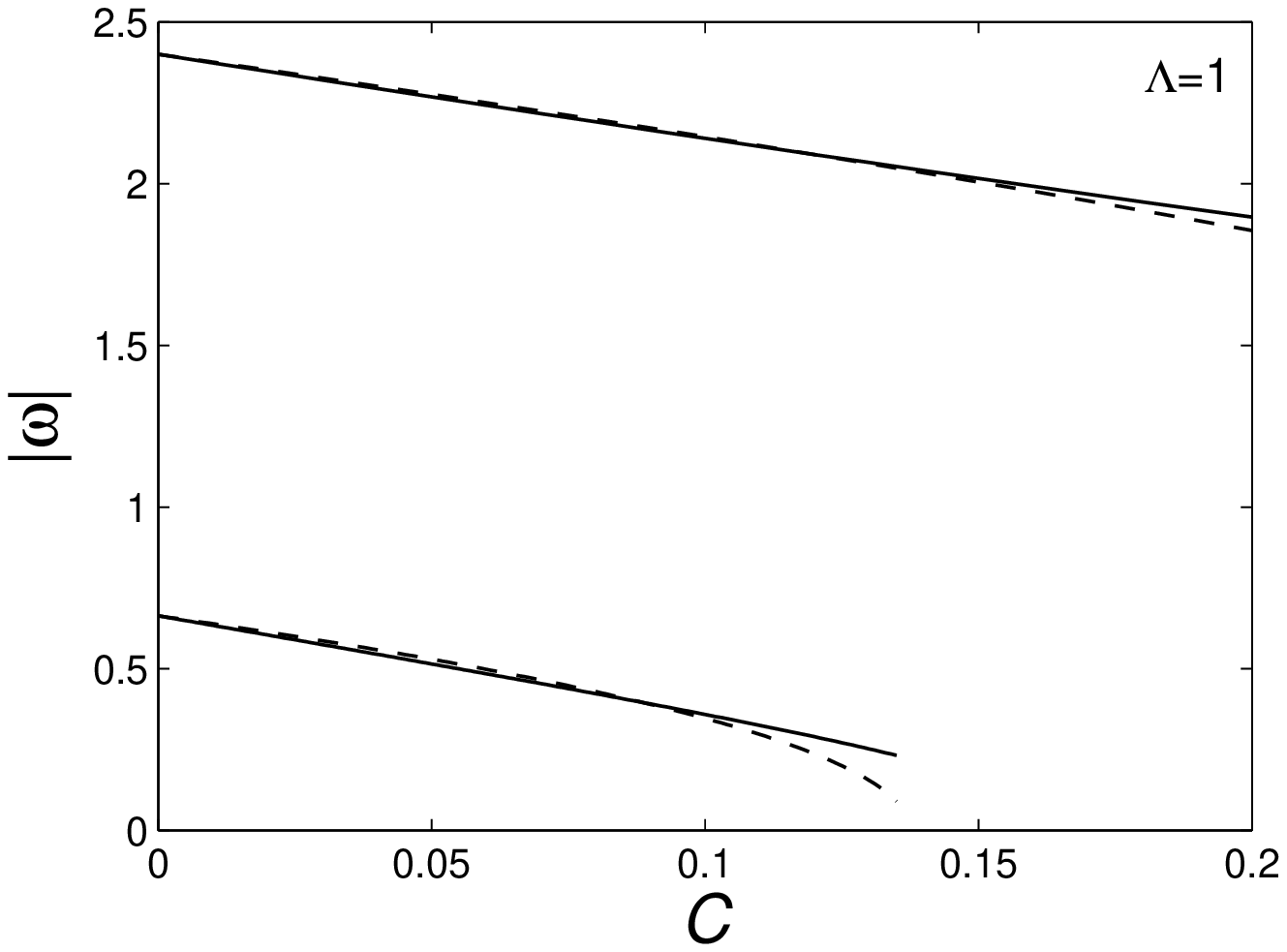}

\caption{Comparisons between the critical eigenvalue for intersite dark
solitons obtained numerically (solid lines) and analytically (dashed
lines) using Eq.~(\ref{eigdn}), for two values of $\gamma$. The upper
curves correspond to $\gamma=0.8$, and the lower ones to $\gamma=0.1$.}
\label{fig1}

\end{figure}

We now compare our analytical results with corresponding numerical
calculations. As for bright solitons, for illustrative purposes we set
$\Lambda=1$.

We start by checking the validity of our analytical approximation for
the critical eigenvalue associated with the phase mode. As explained
above, the change in the position of the eigenvalues relative to the
continuous spectrum at $C=0$ occurs at $\gamma=1/5$. Therefore we
consider the two values $\gamma=0.1$ and $\gamma=0.6$, representing both
cases. Figure~\ref{fig4} depicts a comparison between our analytical
result Eq.~(\ref{eigd}) and the numerical computations, from which we
conclude that the prediction is quite accurate for small $C$. The
accuracy can be improved if one includes further orders in the
perturbative expansion Eq.~(\ref{eigdfurther}), and this improvement is
shown in the same figure by the dotted line.

The eigenvalue structure of onsite dark solitons is depicted in
Fig.~\ref{fig5}; left and right panels refer respectively to conditions
just before and just after a collision resulting in an instability.

As sketched in Fig.~\ref{sketch3}, for $\gamma<1/5$ the instability is
caused by a collision between the eigenvalue and one edge of the
continuous spectrum. On the other hand, when $\gamma\geq1/5$ the
instability is caused by a collision between the eigenvalue and its twin at
the origin (see the bottom panels of Fig.~\ref{fig5}).

We now proceed to evaluate the (in)stability region of this solution in
$(C,\gamma)$ space. Shown in Fig.~\ref{fig7} is again the maximum of the
imaginary part of the eigenvalue, together with our approximation to the
(in)stability boundary. The white dashed line represents
Eq.~(\ref{gamma2}), corresponding to the instability caused by the
collision with the continuous spectrum. Equation~(\ref{gamma1}) is
represented by the white dash-dotted line, which corresponds to the
other instability mechanism. In addition, pink dashed and dash-dotted
lines show, respectively, Eq.~(\ref{gamma2further}) and
Eq.~(\ref{gamma1further}), where a better analytical approximation is
obtained.

An important observation from the figure is that there is an interval of
values of $\gamma$ in which the onsite dark soliton is always stable, for any
value of the coupling constant $C$. This indicates that a parametric
driving can fully suppress the oscillatory instability reported for the
first time in~\cite{joha99}.

\subsubsection{Intersite dark solitons}

\begin{figure*}[tbhp]
\centering
\subfigure[$\gamma=0.1,\,C=0.05$]
{\includegraphics[width=8cm]{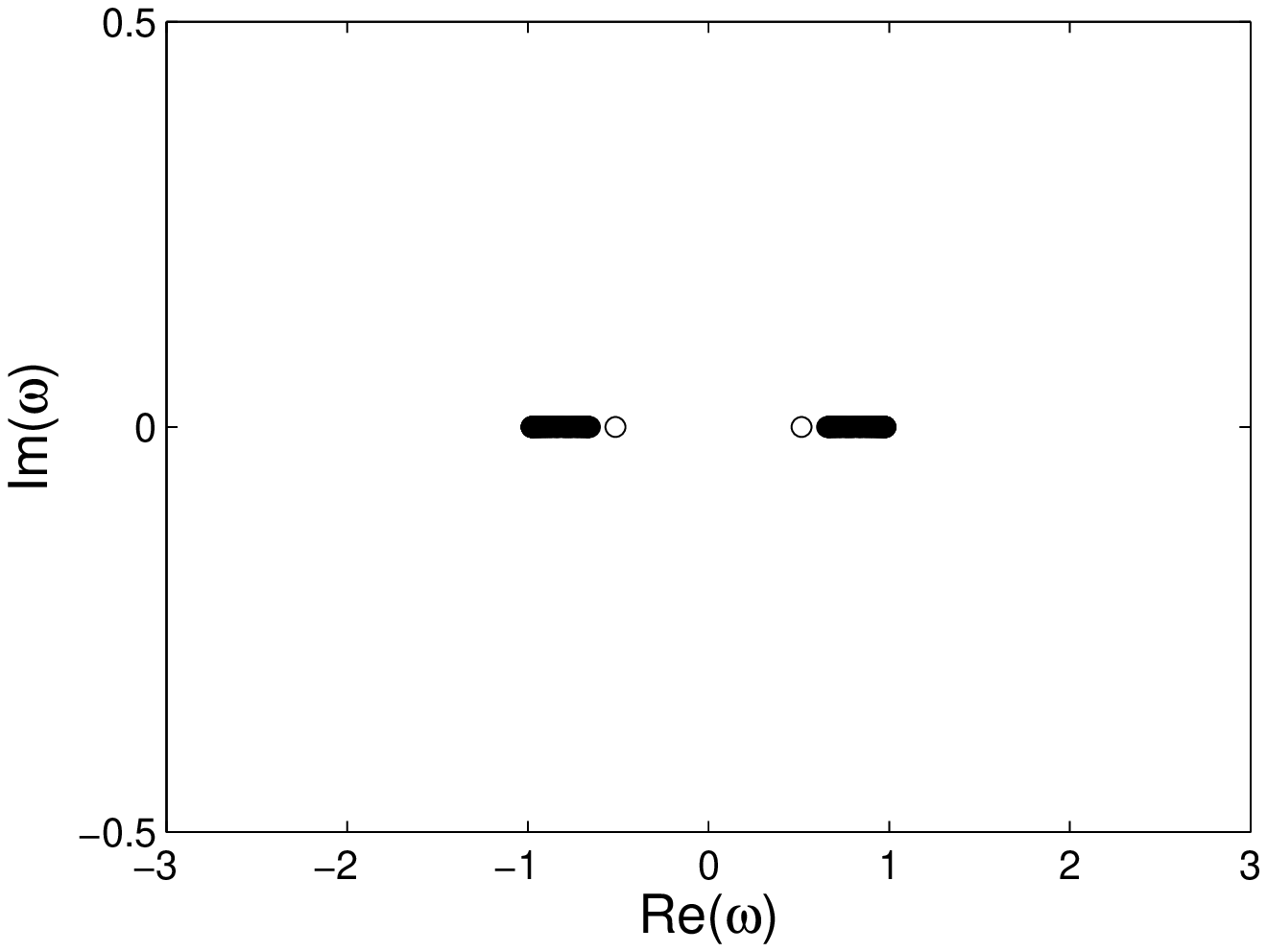}}
\subfigure[$\gamma=0.1,\,C=0.5$]
{\includegraphics[width=8cm]{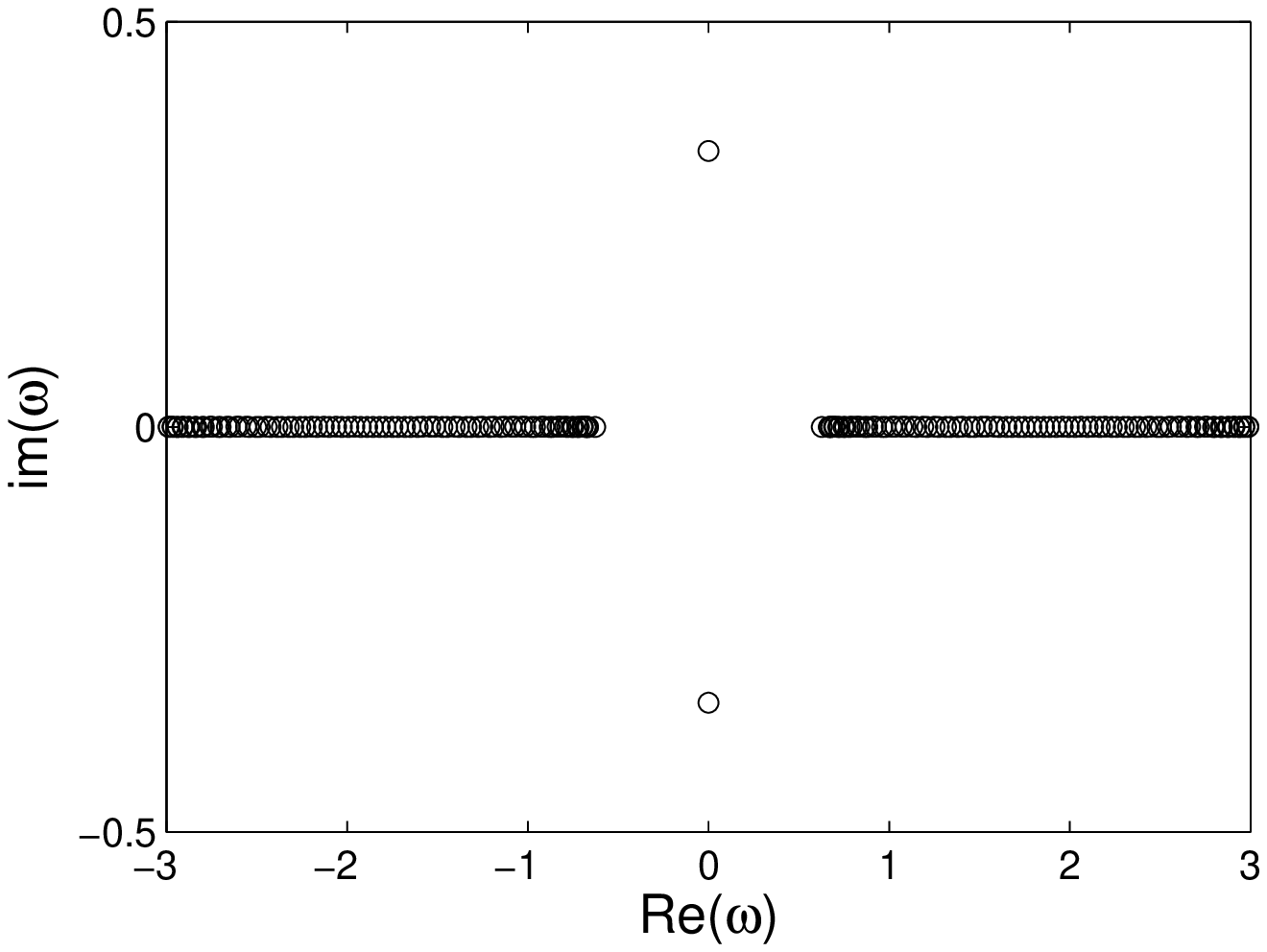}}\\
\subfigure[$\gamma=0.8,\,C=0.5$]
{\includegraphics[width=8cm]{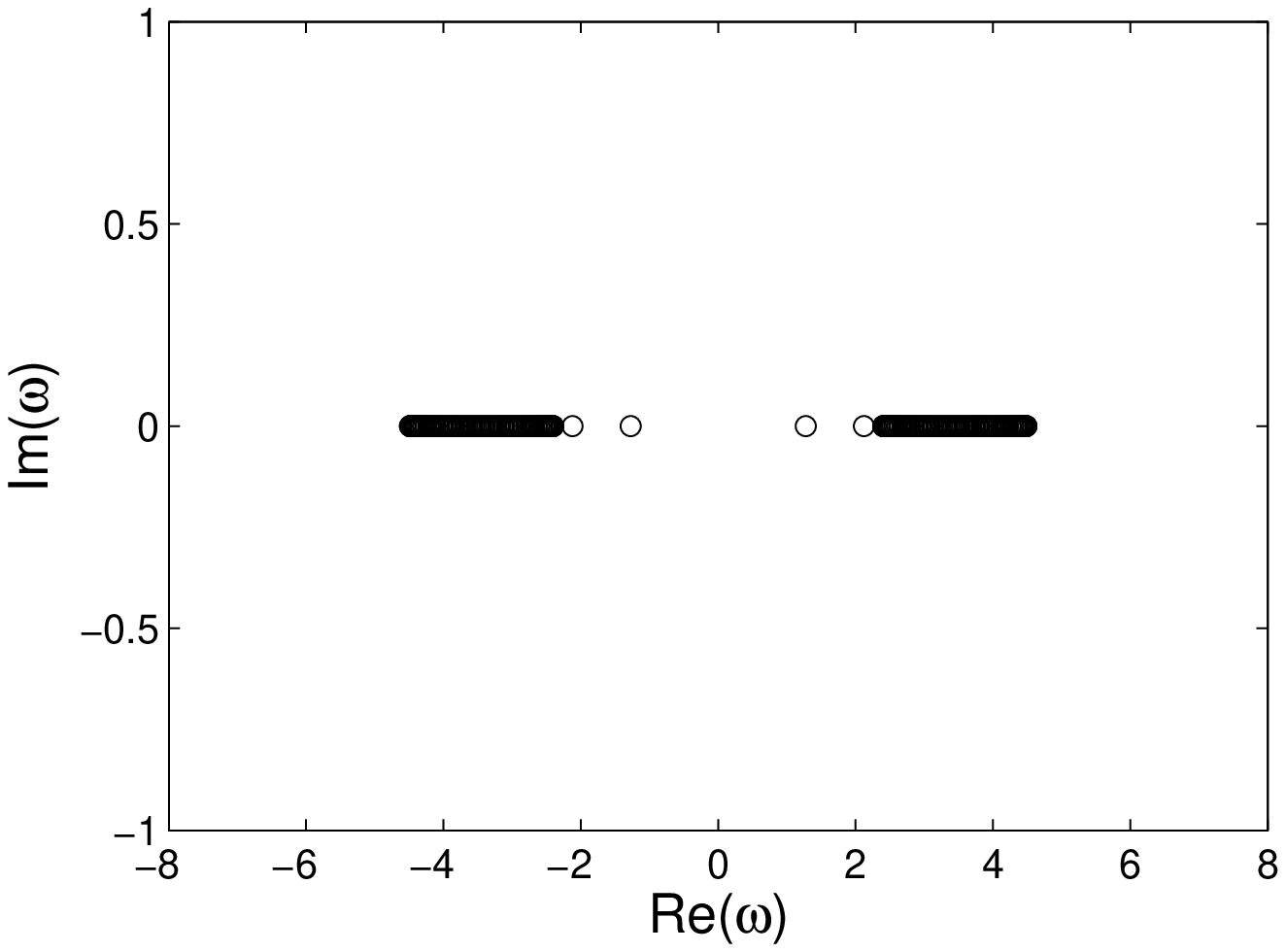}}
\subfigure[$\gamma=0.8,\,C=2$]
{\includegraphics[width=8cm]{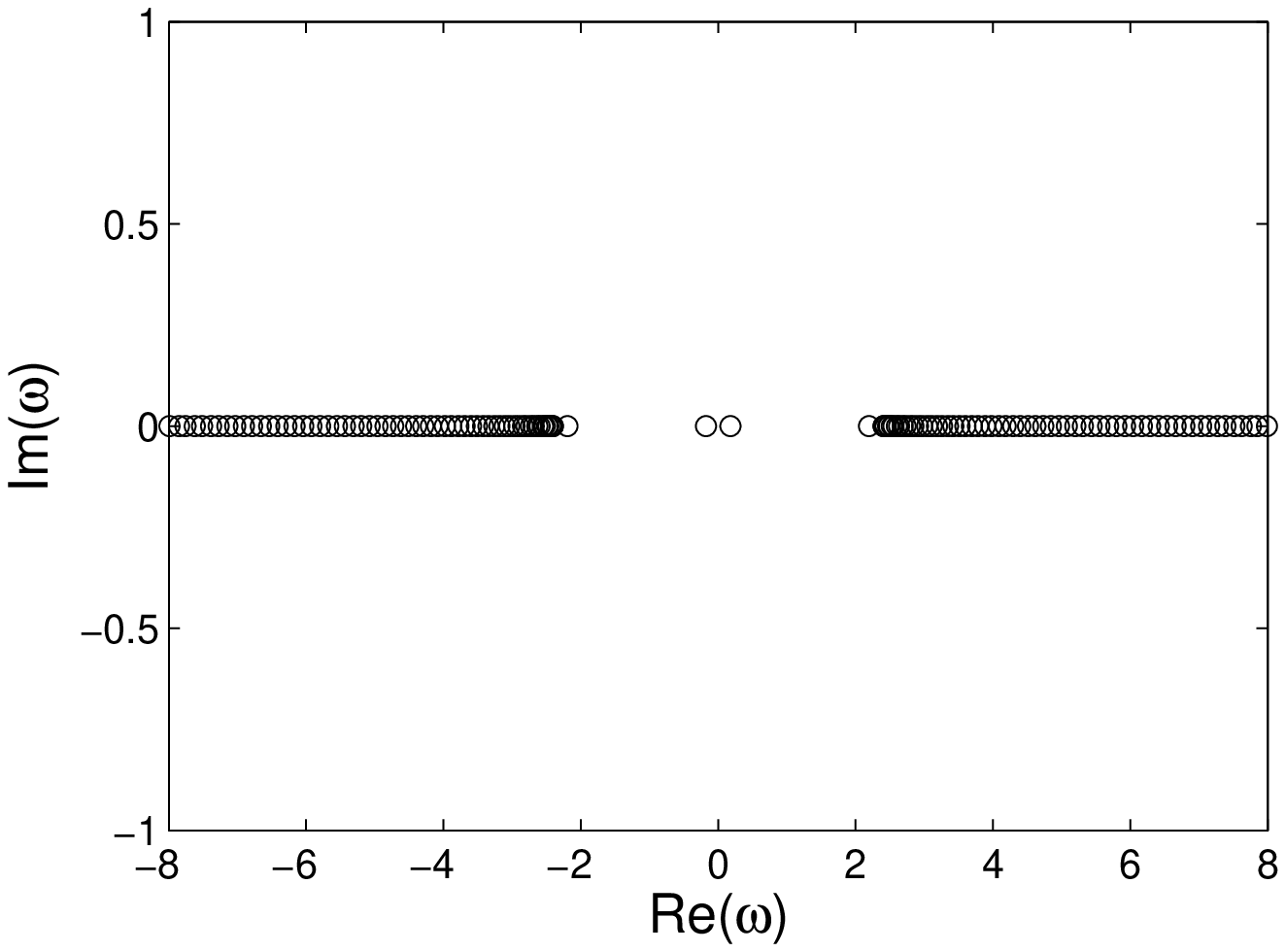}}

\caption{The eigenvalue structure of intersite dark solitons with
parameter values as indicated in the caption for each
panel.}\label{fig2}

\end{figure*}

\begin{figure}[tbhp]
\centering
\includegraphics[width=8cm]{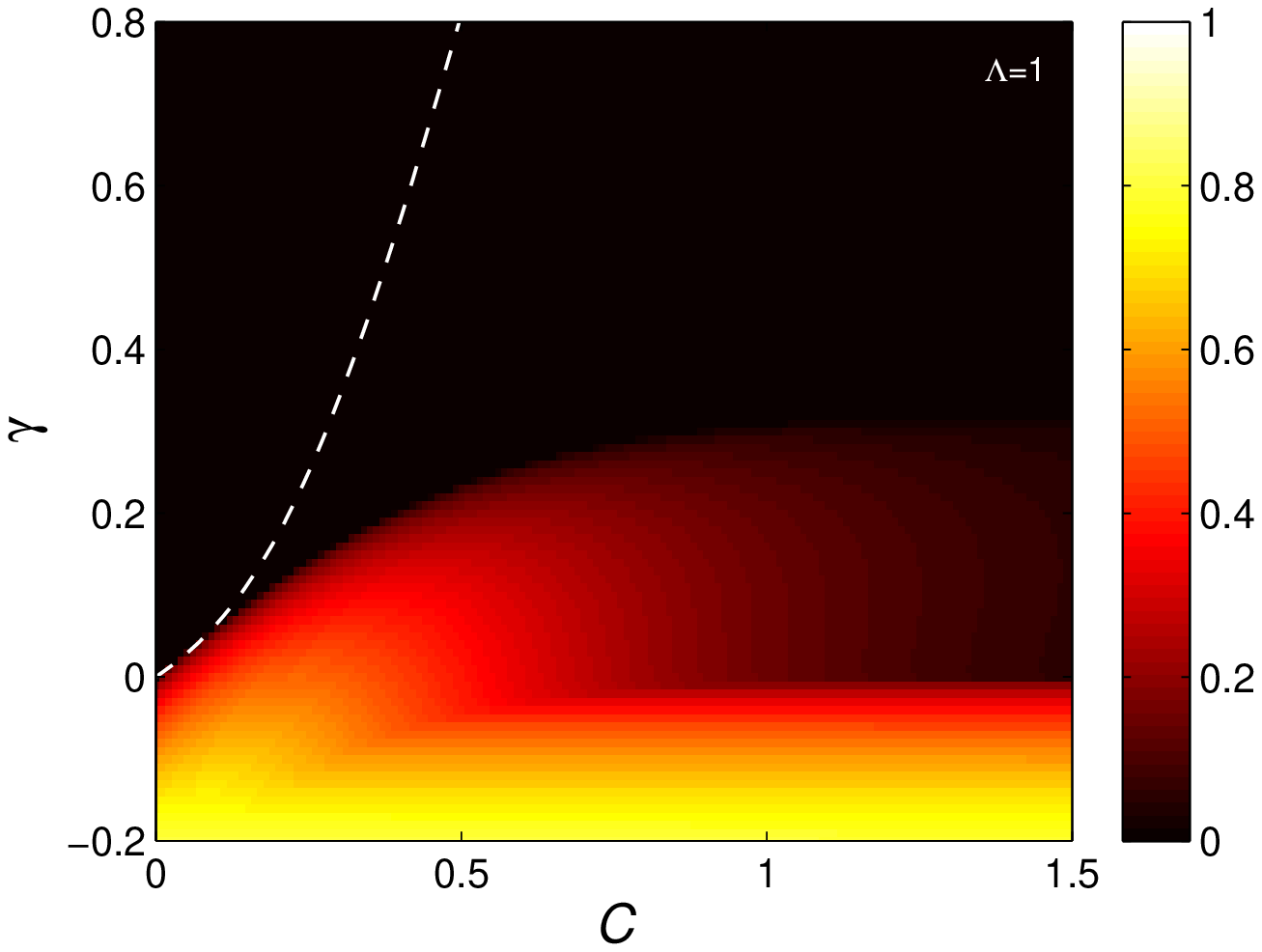}

\caption{(Colour online) As Fig.~\ref{fig7}, but for an intersite dark
soliton. The white dashed line is our analytical approximation
Eq.~(\ref{gammacr}).} \label{fig3}

\end{figure}

Now we examine intersite dark solitons.

Firstly, Fig.~\ref{fig1} shows the analytical prediction for the
critical eigenvalue, given by Eq.~(\ref{eigdn}), compared to numerical
results. We see that the approximation is excellent for small $C$ and
its range of validity is wider for larger values of $\gamma$. The
eigenvalue structure of this configuration is shown in Fig.~\ref{fig2}
for the two values $\gamma=0.1,0.8$. The mechanism of instability
explained in the section above can be seen clearly in the top panels of
Fig.~\ref{fig2}.

It is interesting to note that a parametric driving can also fully
suppress the oscillatory instability of an intersite dark soliton. As
shown in the bottom panels of Fig.~\ref{fig2}, there are values of the
parameter $\gamma$ for which no instability-inducing collision ever
occurs. The (in)stability region of this configuration is summarized in
Fig.~\ref{fig3}, where we see that for any $C$ and $\gamma>0.3$ an
intersite dark soliton is always stable. Our analytical prediction for
the onset of instability is given by the dashed line in that figure. We
observe that for relatively small $C$, the prediction of
Eq.~(\ref{eigdn}) is reasonably close to the numerical results.

\section{Discussion}

In the sections above we discussed the existence and the stability of
localized modes through our reduced DNLS equation (\ref{gov}). In this
section, we confirm the relevance of our findings through solving
numerically the original time-dependent equation (\ref{eq1}). We use a
Runge--Kutta integration method, with the initial condition
$\varphi_n=2\epsilon u_n$ and $\dot\varphi_n=0,$ where $u_n$ is the
static solution of the DNLS (\ref{gov}) and $\epsilon$ is the small
parameter of Sec.~I. Throughout this section, we use the illustrative
value $\epsilon=0.2$.

\begin{figure*}[tbhp]
\centering
\includegraphics[width=8cm]{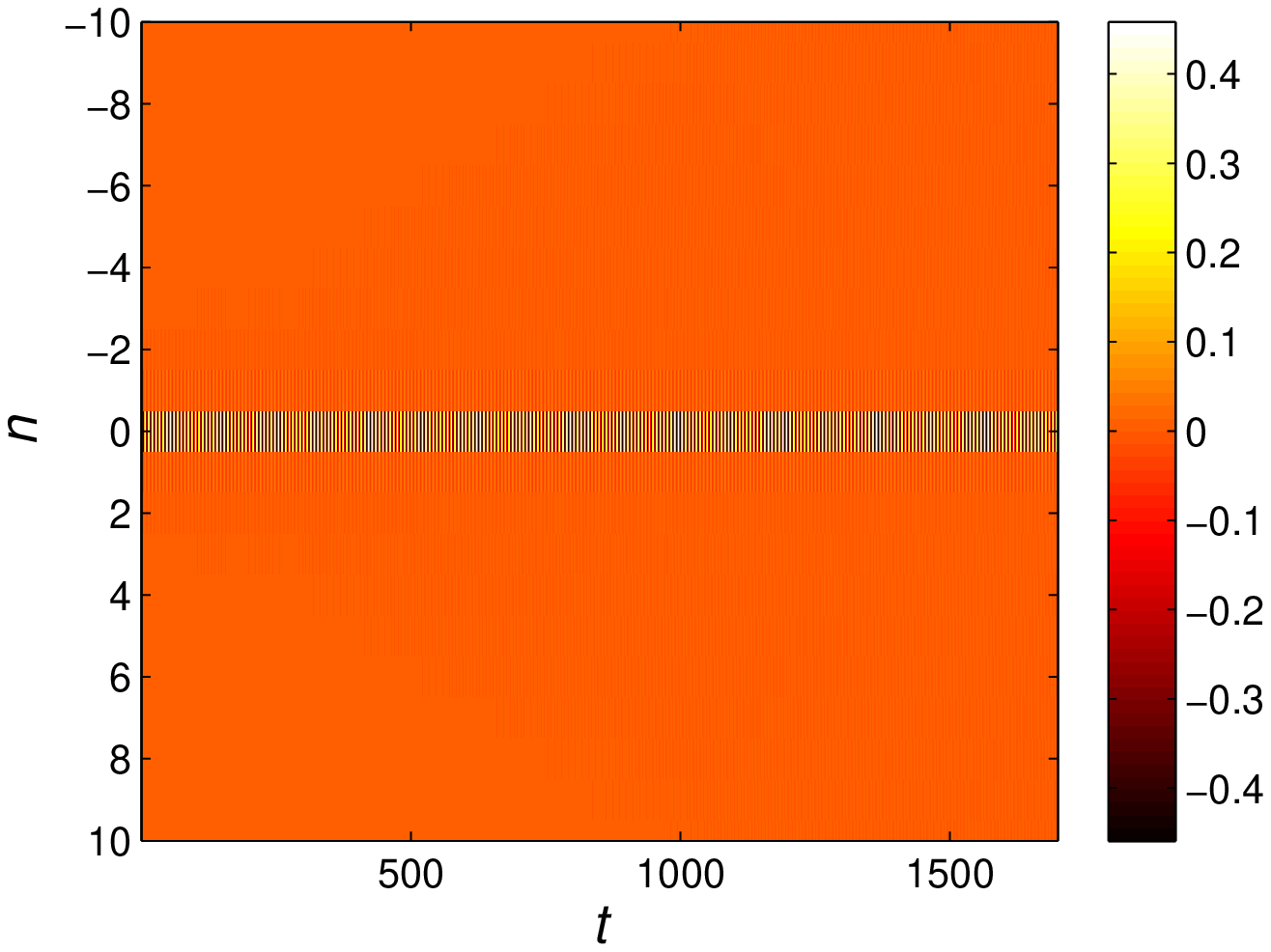}
\includegraphics[width=8cm]{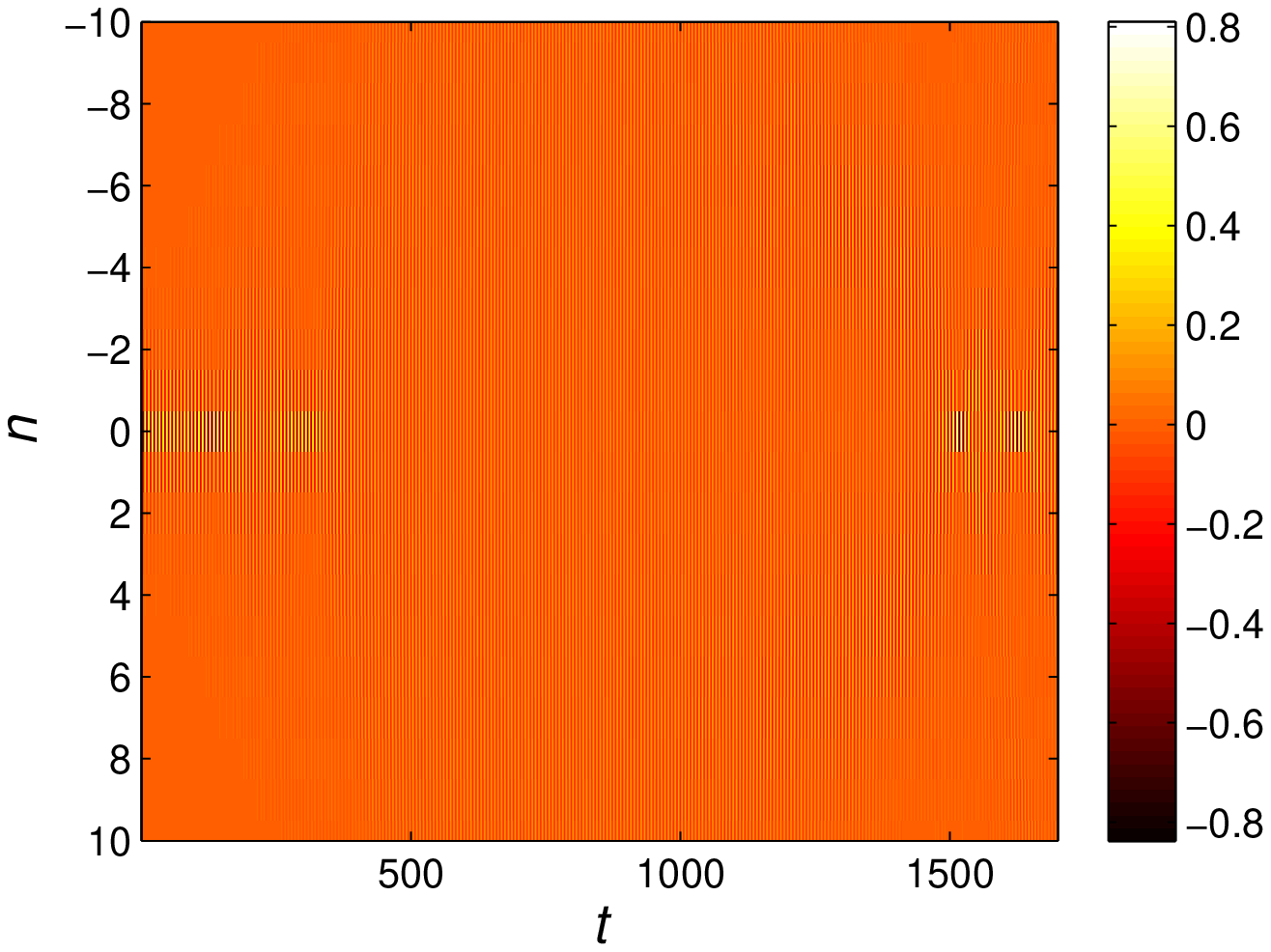}

\caption{(Colour online) The spatio-temporal evolution of an onsite
bright soliton governed by the original time-dependent parametrically
driven Klein--Gordon system (\ref{eq1}), with $\epsilon=0.2$ and
$\gamma=0.1$. The left and right panels show a stable and unstable
soliton, at $C=0.1$ and $C=1$, respectively.}\label{figbrightSTorg}

\end{figure*}

\begin{figure*}[tbhp]
\centering
\subfigure[$\gamma=0.1,\,C=0.05$]{\includegraphics[width=8cm]
{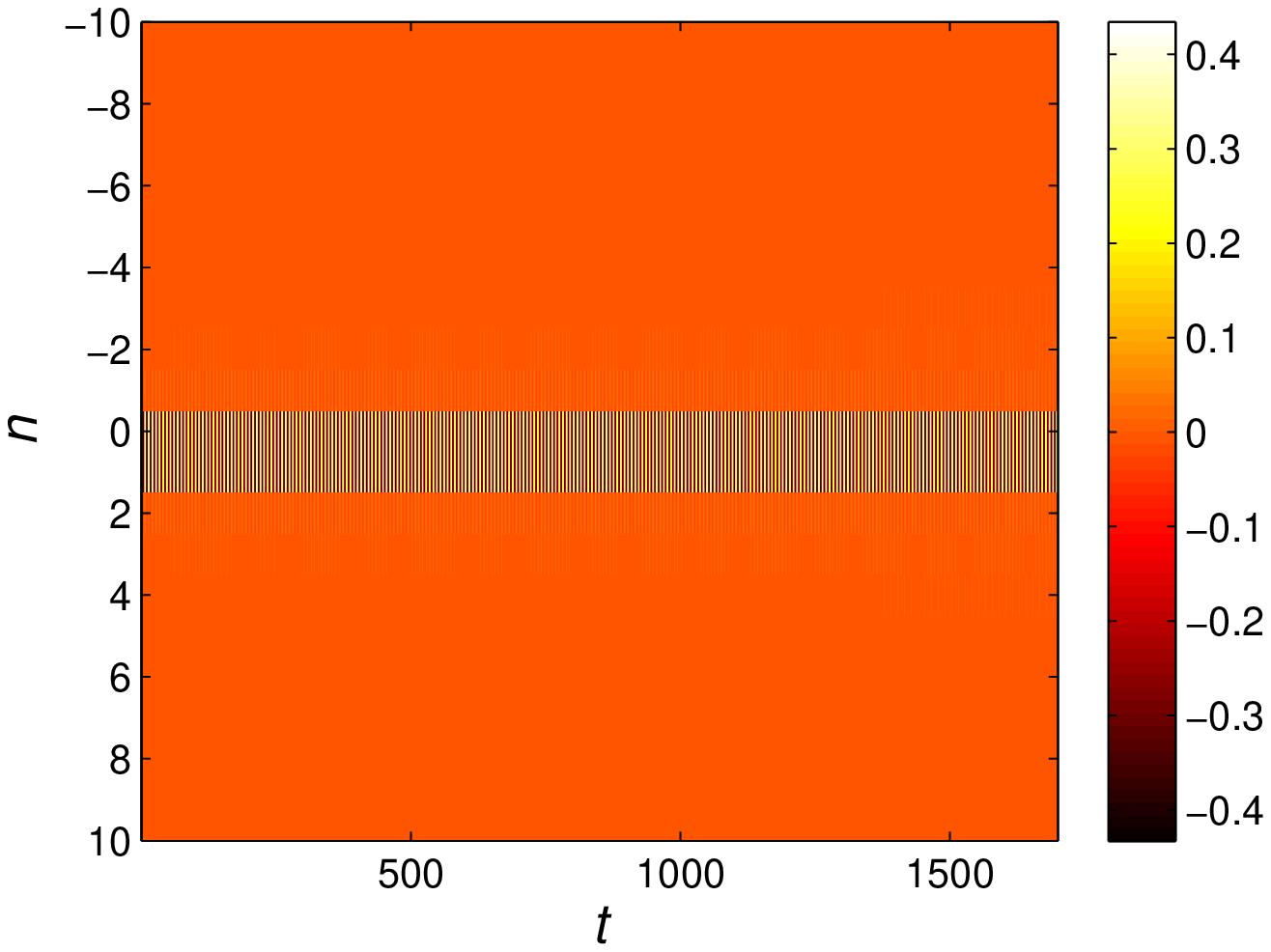}}
\subfigure[$\gamma=0.1,\,C=0.3$]{\includegraphics[width=8cm]
{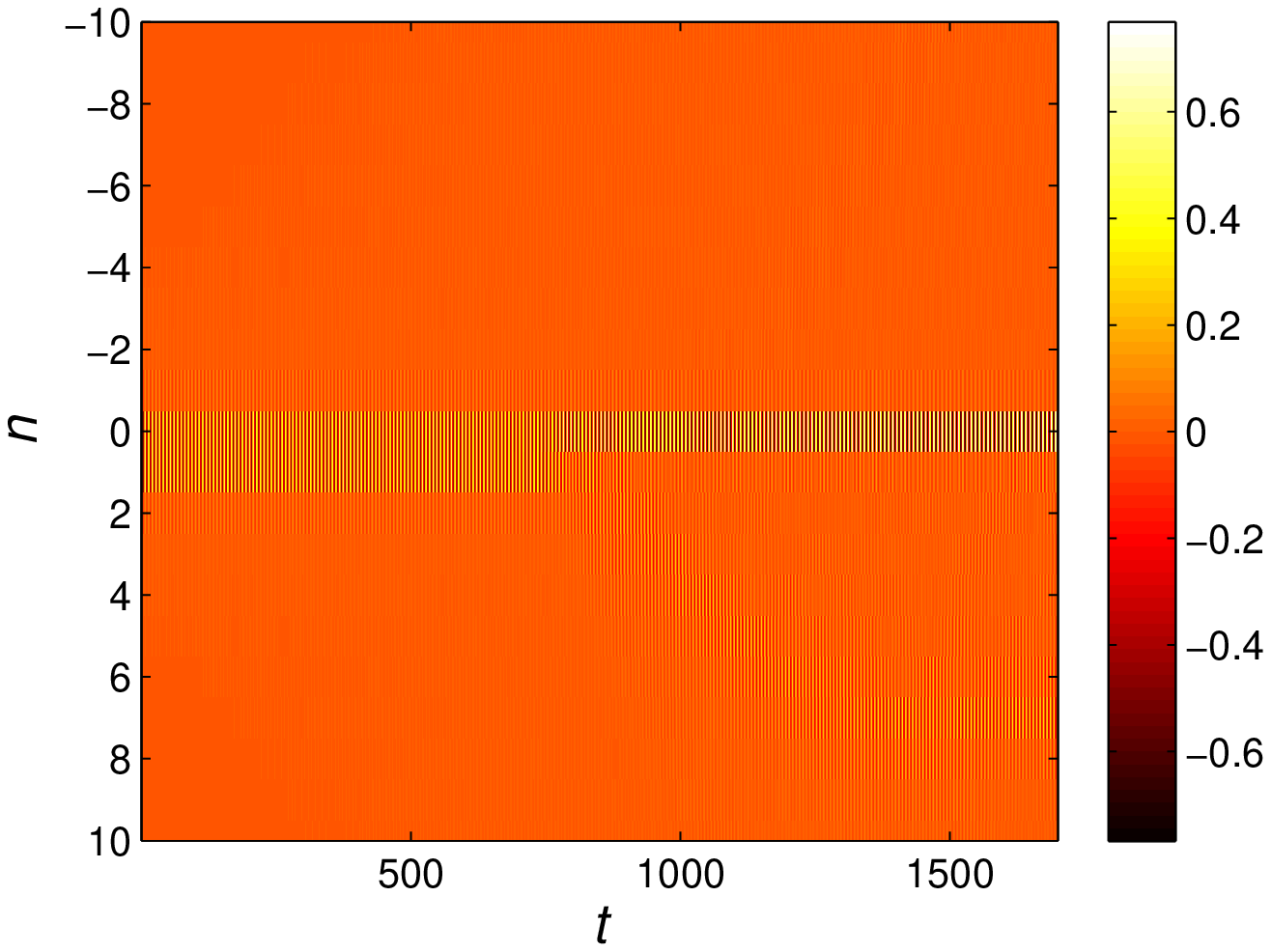}}\\
\subfigure[$\gamma=0.18,\,C=0.05$]{\includegraphics[width=8cm]
{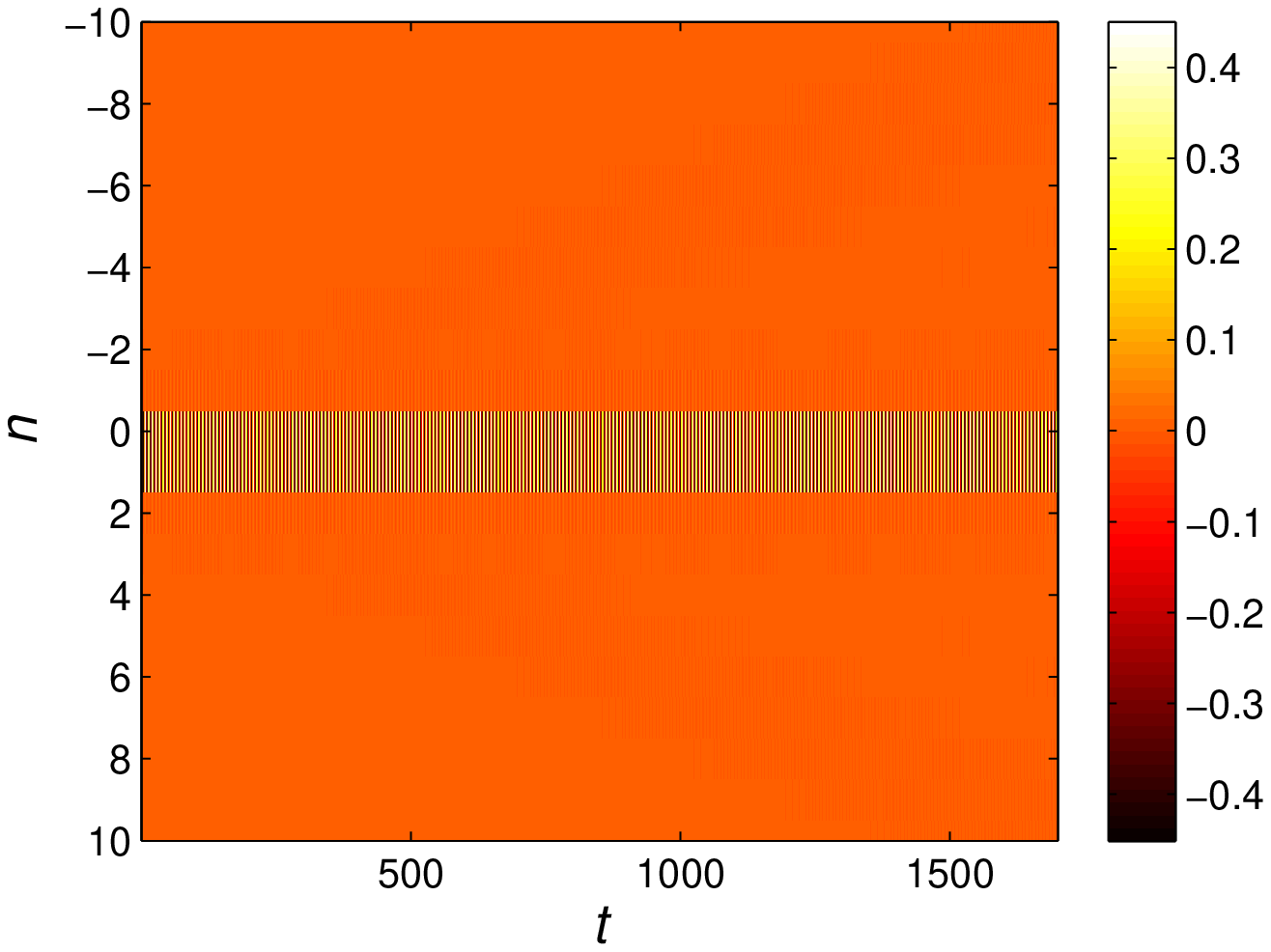}}
\subfigure[$\gamma=0.18,\,C=0.18$]{\includegraphics[width=8cm]
{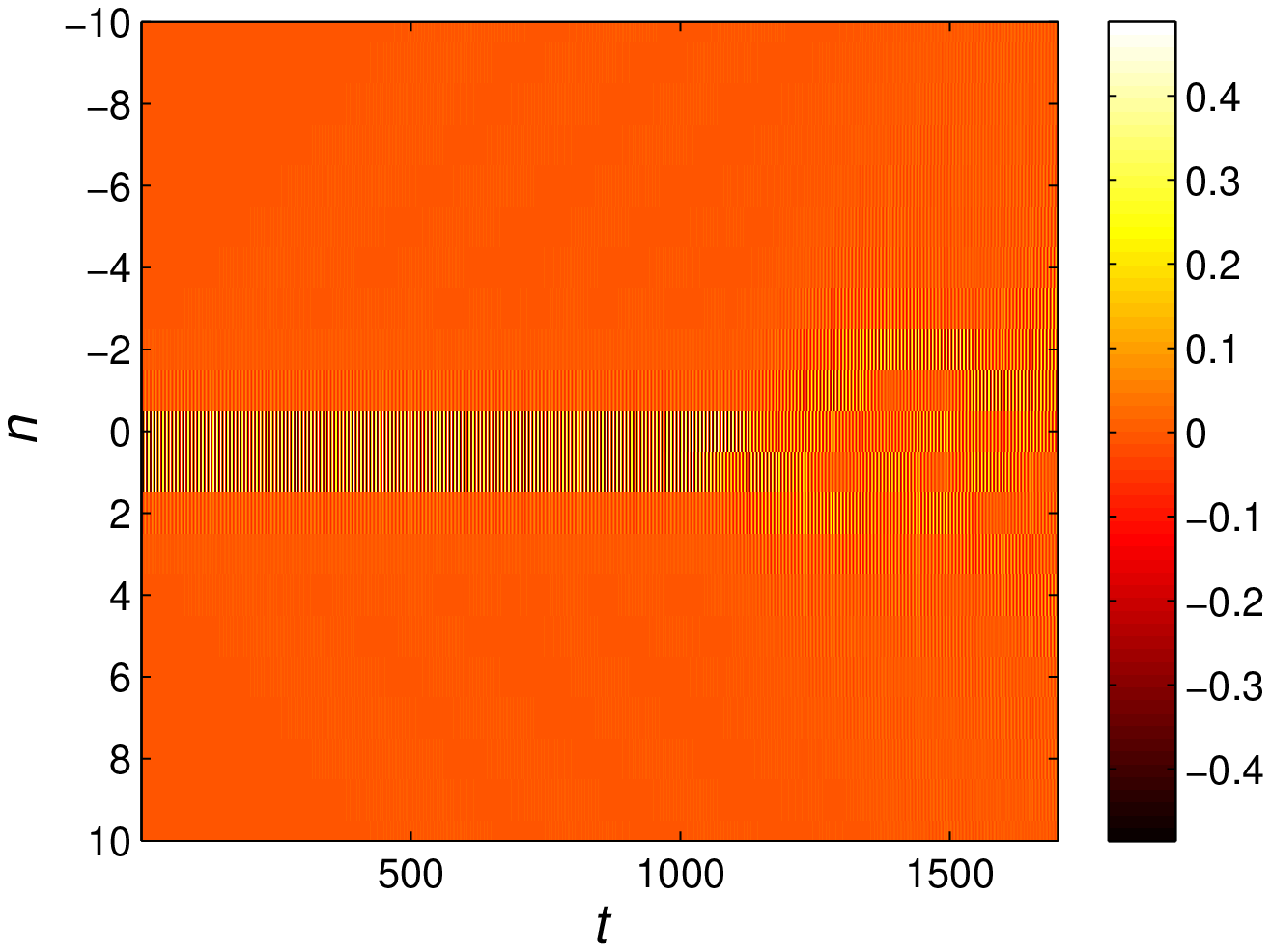}}\\
\subfigure[$\gamma=0.5,\,C=0.05$]{\includegraphics[width=8cm]
{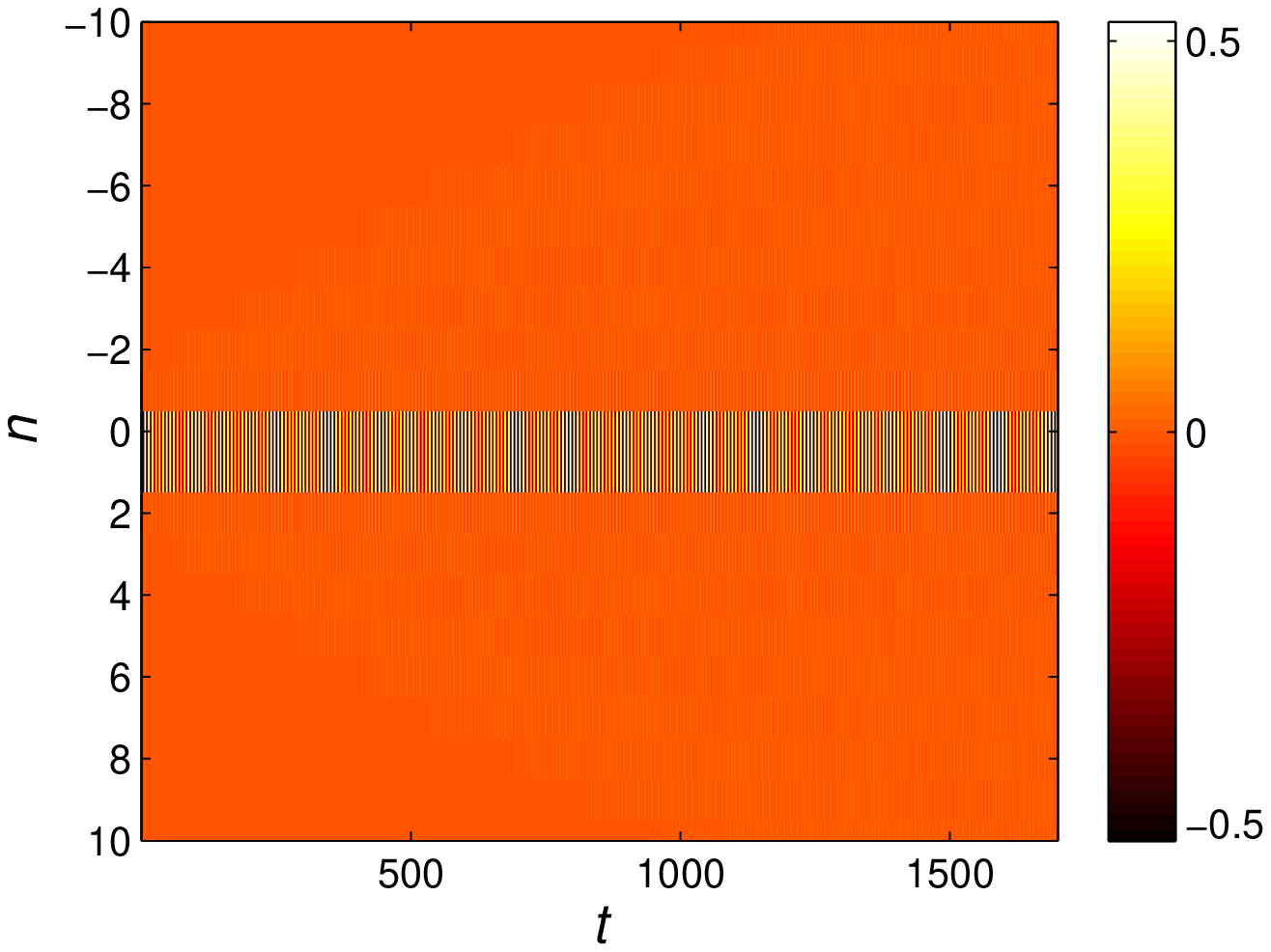}}
\subfigure[$\gamma=0.5,\,C=0.2$]{\includegraphics[width=8cm]
{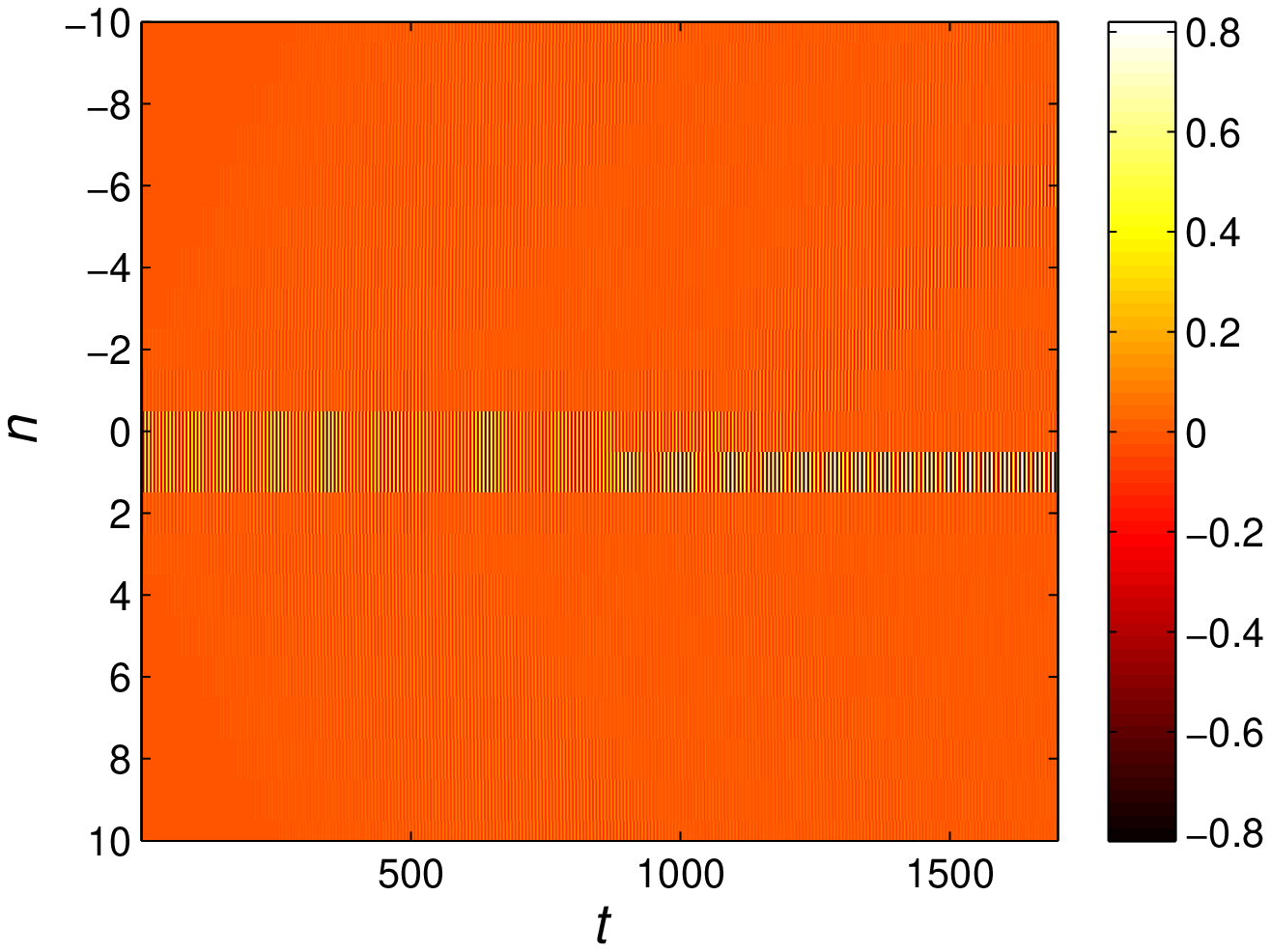}}

\caption{(Colour online) As Fig.~\ref{figbrightSTorg}, but for an
intersite bright soliton, with parameter values as indicated in the
caption for each panel. The initial profile in each panel corresponds to
the same parameters as in Fig.~\ref{figbright3}.}\label{figbrightPorg}

\end{figure*}

\begin{figure*}[tbhp]
\centering
\subfigure[$\gamma=0.1,\,C=0.02$]
{\includegraphics[width=8cm]{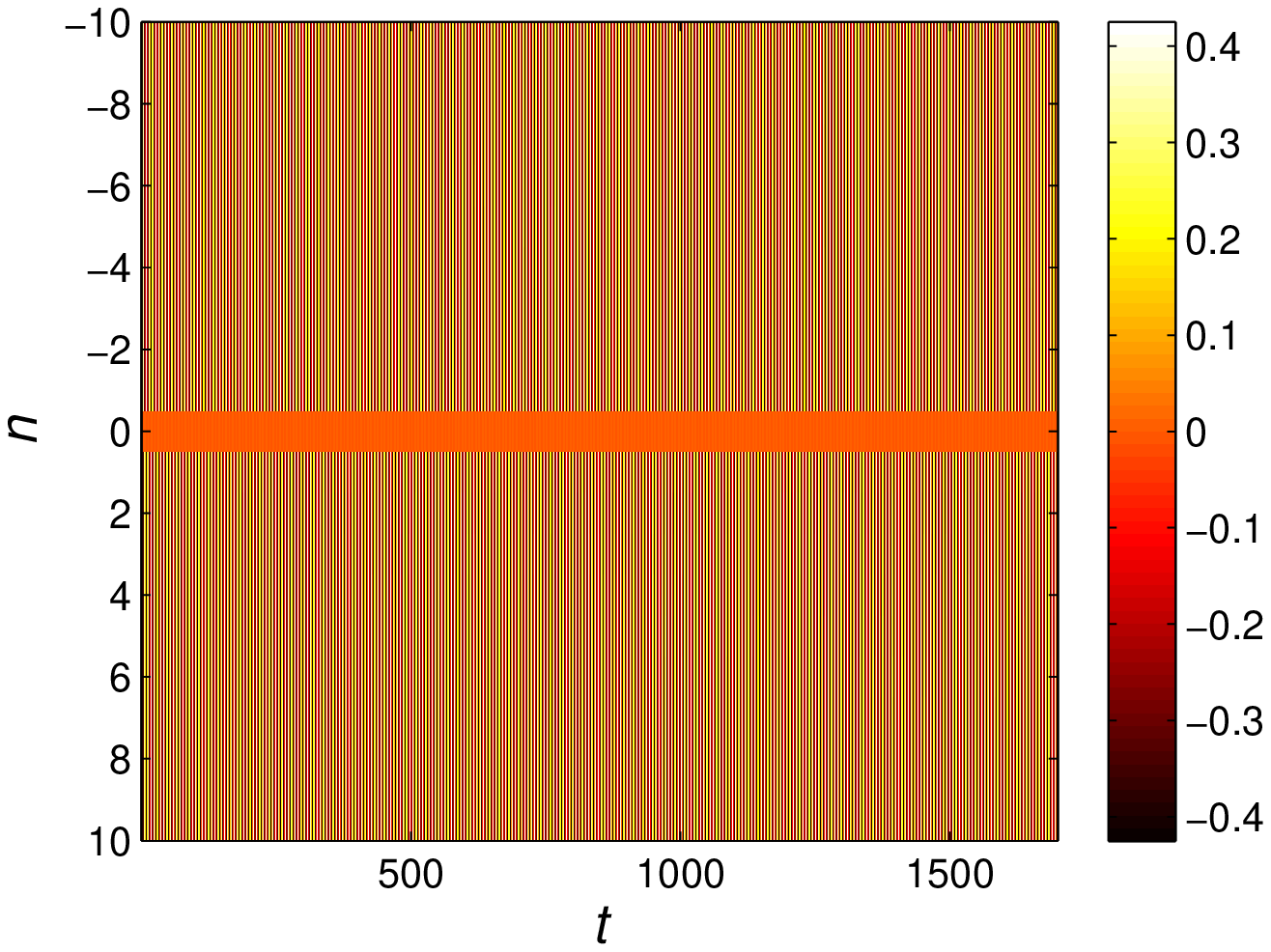}}
\subfigure[$\gamma=0.1,\,C=0.2$]
{\includegraphics[width=8cm]{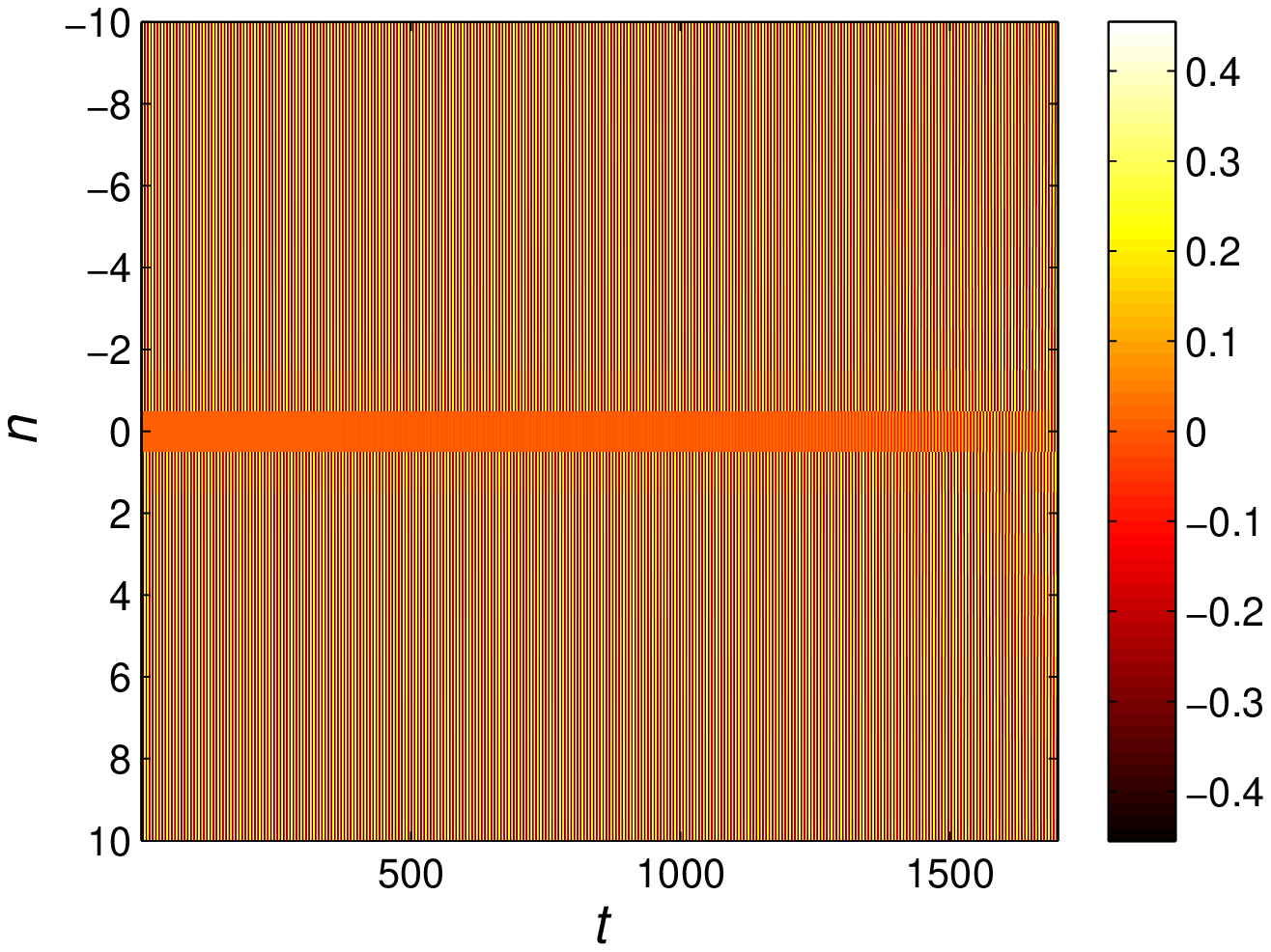}}\\
\subfigure[$\gamma=0.6,\,C=0.01$]
{\includegraphics[width=8cm]{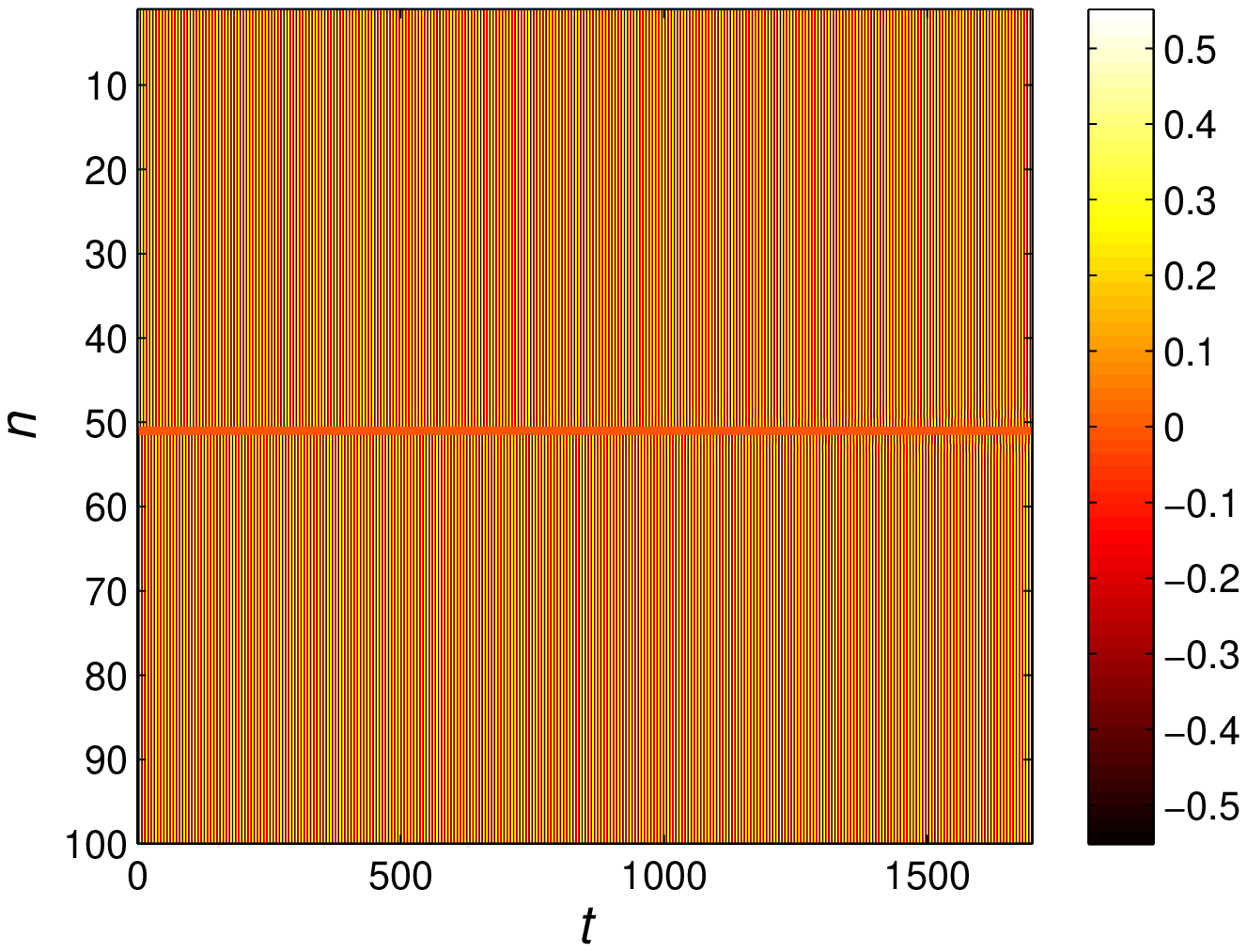}}
\subfigure[$\gamma=0.6,\,C=1$]
{\includegraphics[width=8cm]{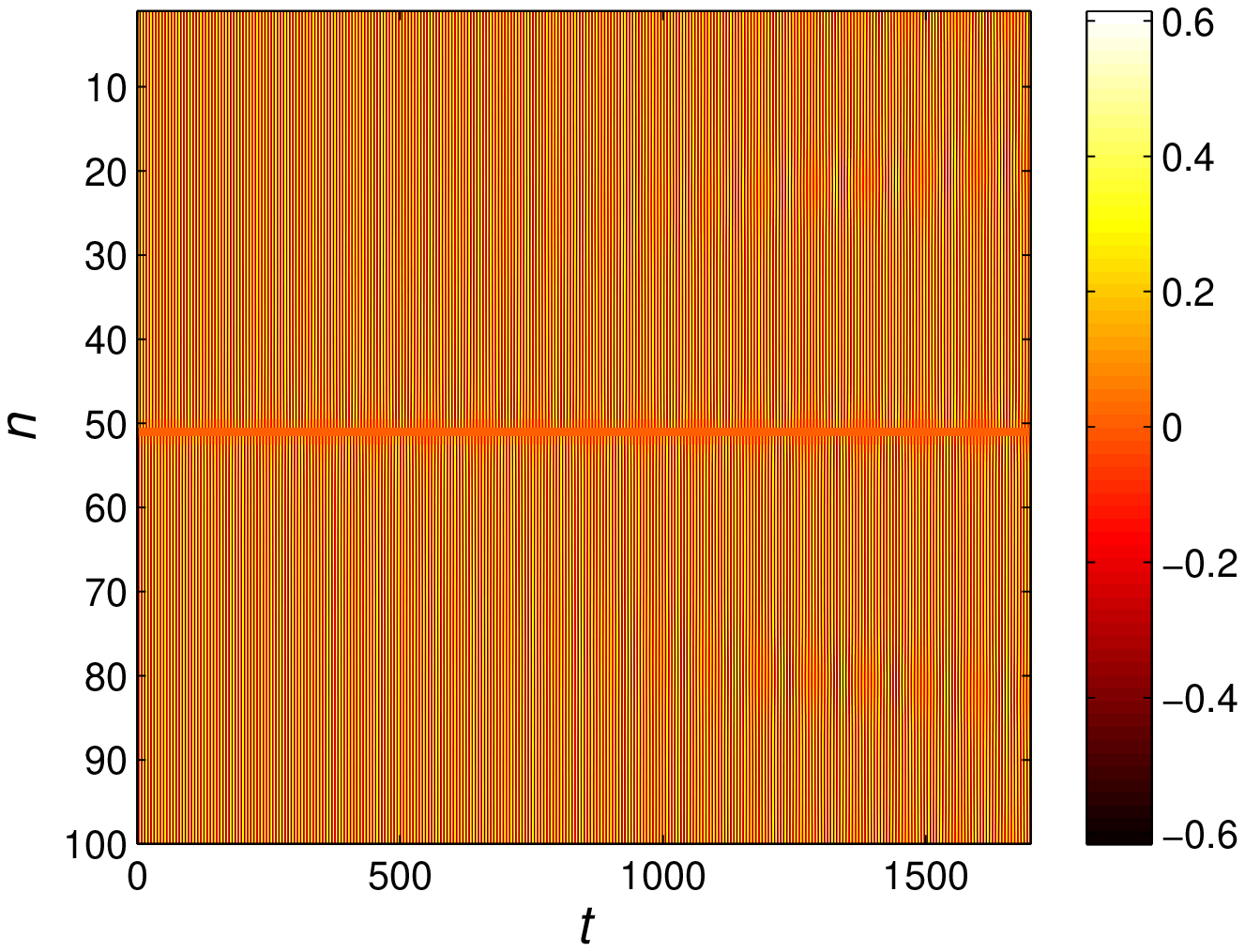}}

\caption{(Colour online) As Fig.~\ref{figbrightSTorg}, but for on-site
dark solitons. The parameter values are as in
Fig.~\ref{fig5}.}\label{evoondark1}

\end{figure*}

\begin{figure*}[tbhp]
\centering
\includegraphics[width=8cm]{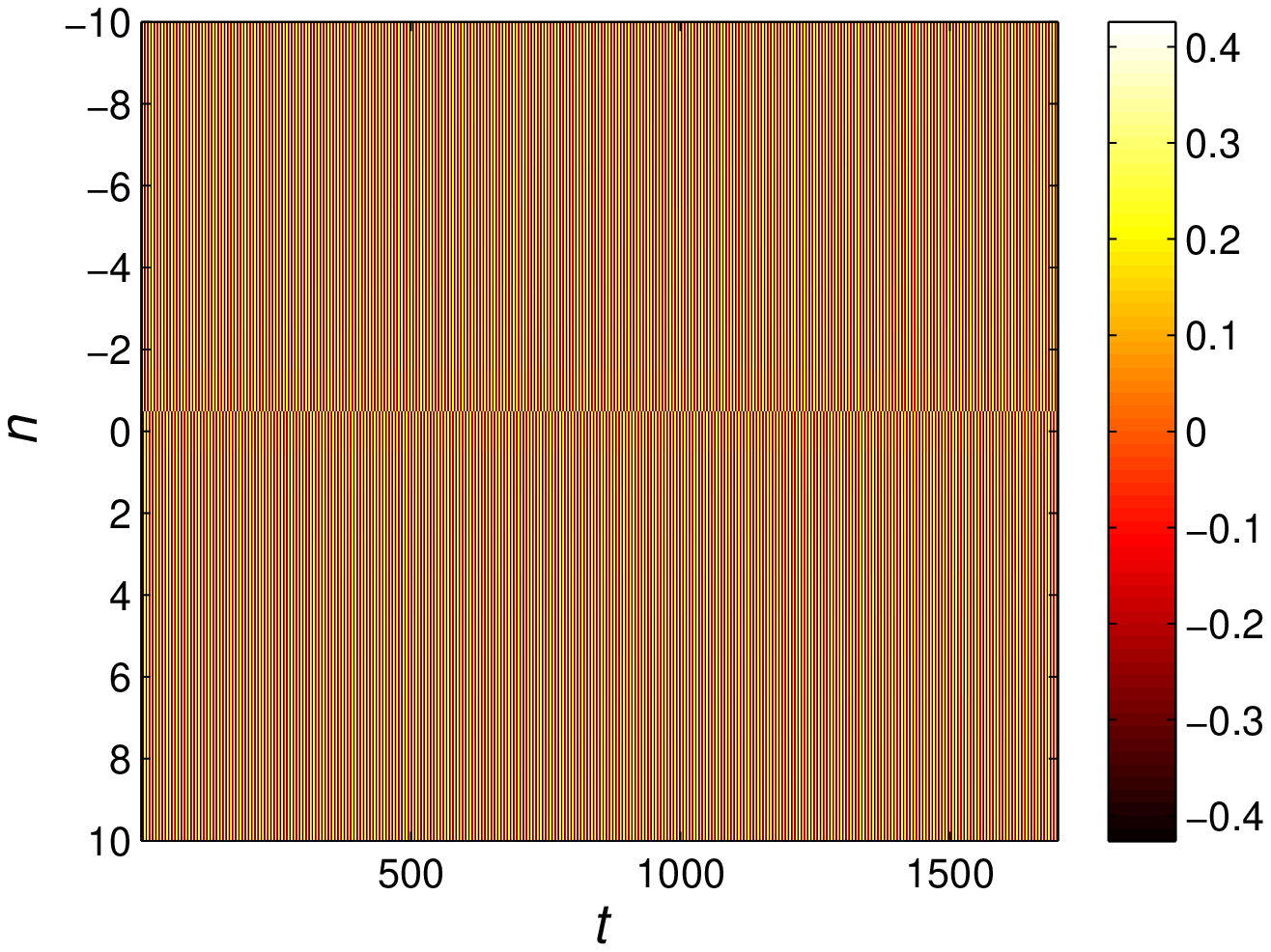}
\includegraphics[width=8cm]{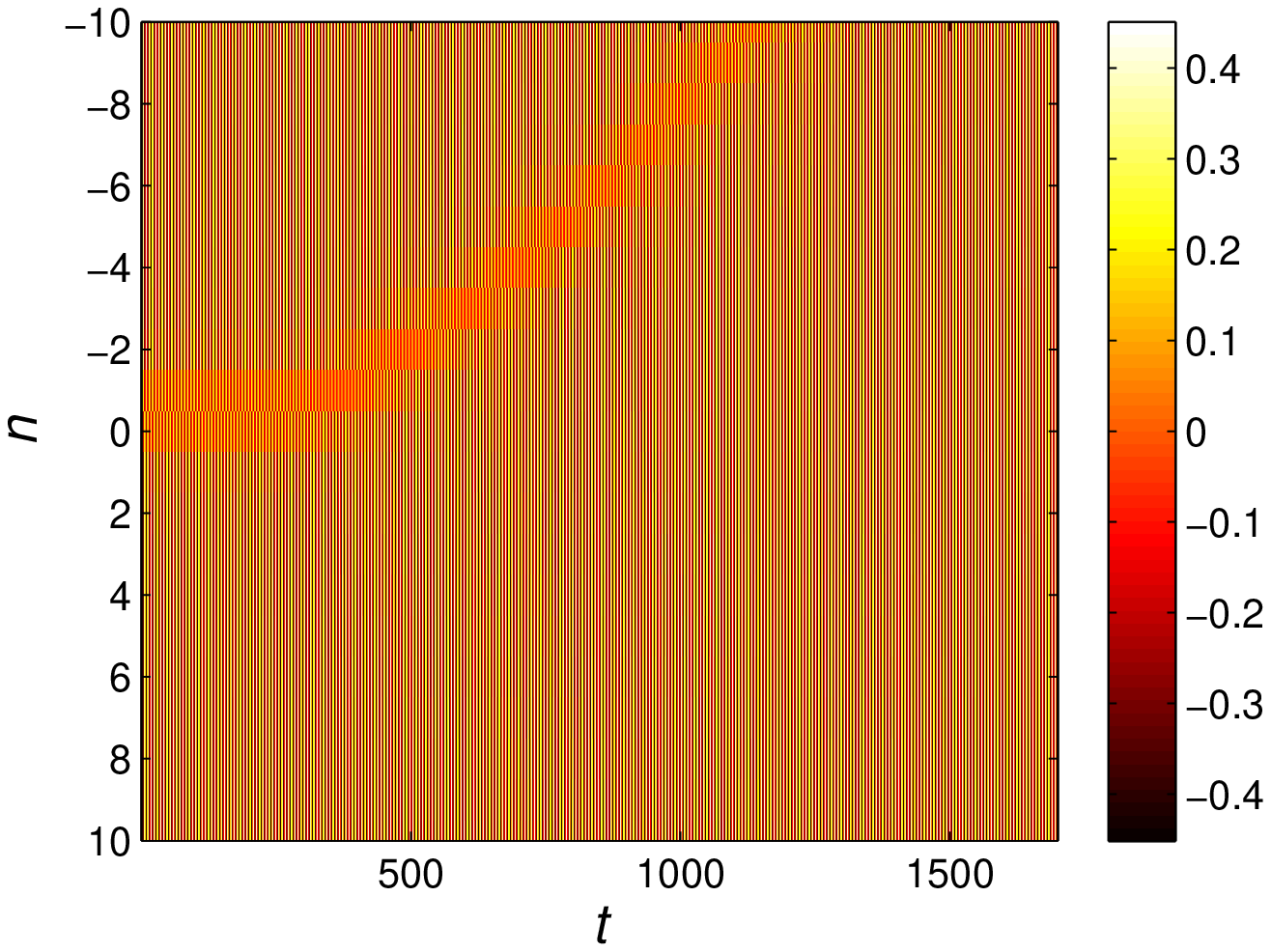}

\caption{(Colour online) As Fig.~\ref{figbrightSTorg}, but for an
intersite dark soliton with $\gamma=0.1$. The left panel shows the
evolution of a stable dark soliton with $C=0.05$, while the right panel
shows the evolution of an unstable dark soliton with
$C=0.5$.}\label{evointdark}

\end{figure*}

Shown in the left and right panels of Fig.~\ref{figbrightSTorg} are the
numerical evolution of a stable and unstable onsite bright soliton,
respectively. From the right panel of the figure, we note that a
parametric driving seems to destroy an unstable soliton. This
observation is similar to the corresponding observation for the dynamics
of an unstable soliton in the DNLS equation (\ref{gov}) reported in
Ref.~\onlinecite{susa06}.

In Fig.~\ref{figbrightPorg} we present the numerical evolution of
intersite bright solitons for the same parameter values as those in
Fig.~\ref{figbright3}, corresponding to each of the instability
scenarios. From the panels in this figure, we see that the typical
dynamics of the instability is in the form of soliton destruction or
discharge of a traveling breather.

We have also examined the dynamics of onsite dark solitons in the
Klein--Gordon system (\ref{eq1}). Shown in Fig.~\ref{evoondark1} is
the numerical evolution of a solution with the eigenvalue structure
illustrated in Fig.~\ref{fig5}. The instability of an unstable onsite
dark soliton typically manifests itself in the form of oscillations in
the location of the soliton center about its initial position (top right
panel) or oscillations in the width of the soliton (bottom right panel).

Finally, we illustrate the dynamical behavior of an unstable intersite
dark soliton in Fig.~\ref{evointdark}, from which we see that the
instability makes the soliton travel. This dynamics is similar to that
reported in Ref.~\onlinecite{fitr07}.

\section{Conclusion}

In this paper, we have considered a parametrically driven Klein--Gordon
system describing nanoelectromechanical systems. Using a multiscale
expansion method we have reduced the system to a parametrically driven
discrete nonlinear Schr\"odinger equation. Analytical and numerical
calculations have been performed to determine the existence and
stability of fundamental bright and dark discrete solitons in the
Klein--Gordon system through use of the Schr\"odinger equation. We have
shown that the presence of a parametric driving can destabilize an
onsite bright soliton. On the other hand, a parametric driving has also
been shown to stabilize intersite bright and dark discrete solitons. We
even found an interval in $\gamma$ for which a discrete dark soliton is
stable for any value of the coupling constant, i.e.\ a parametric
driving can suppress oscillatory instabilities. Stability windows for
all the fundamental solitons have been presented and approximations
using perturbation theory have been derived to accompany the numerical
results. Numerical integrations of the original Klein--Gordon system have
demonstrated that our analytical and numerical investigations of the
discrete nonlinear Schr\"odinger equation provide a useful guide to
behavior in the original system.


\end{document}